\documentclass[nopubldata]{lpor2014}

\begin{document}

\titlefigure{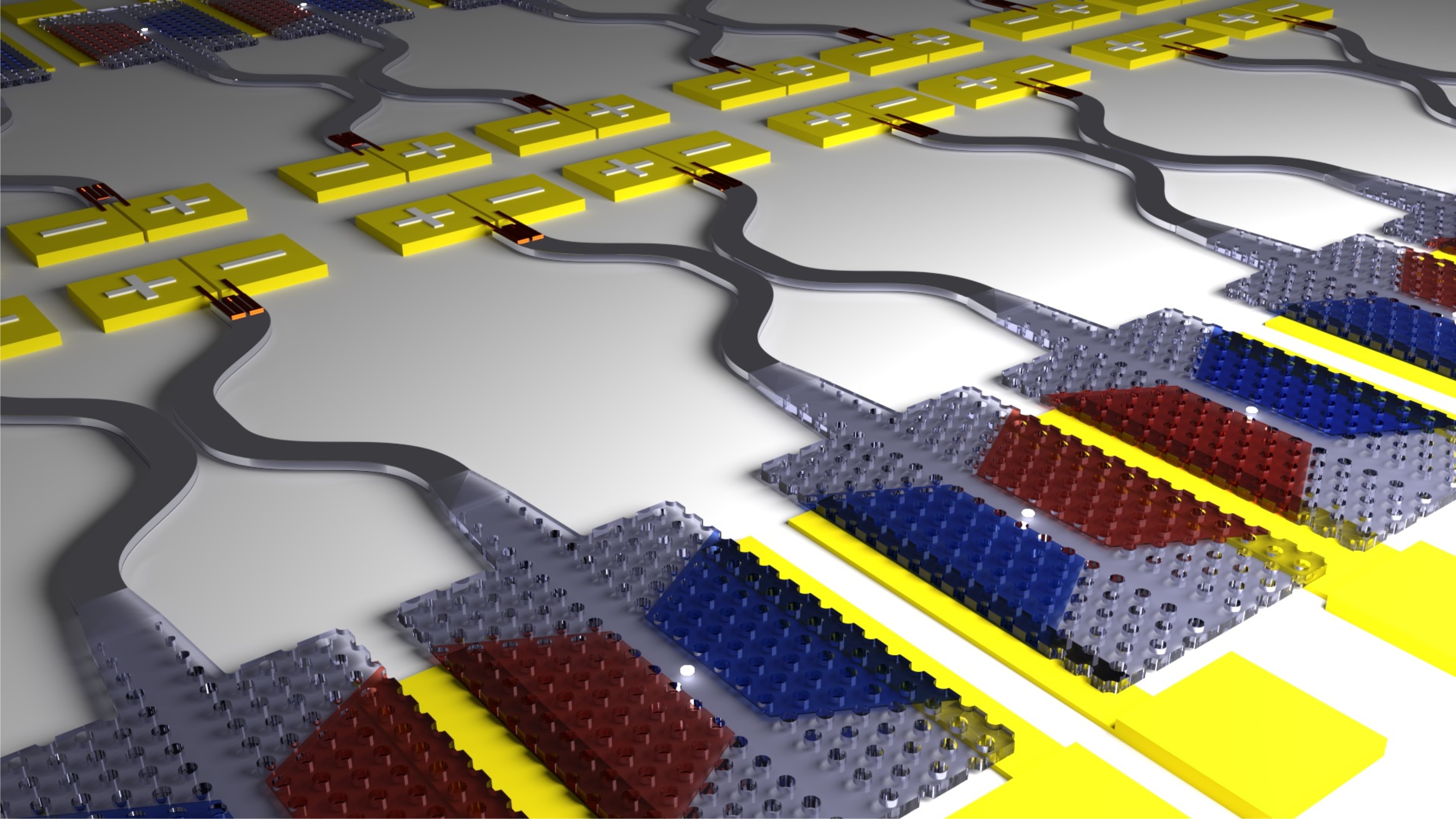}
\abstract{The recent progress in integrated quantum optics has set the stage for the development of an integrated platform for quantum information processing with photons, with potential applications in quantum simulation. Among the different material platforms being investigated, direct-bandgap semiconductors and particularly gallium arsenide (GaAs) offer the widest range of functionalities, including single- and entangled-photon generation by radiative recombination, low-loss routing, electro-optic modulation and single-photon detection. This paper reviews the recent progress in the development of the key building blocks for GaAs quantum photonics and the perspectives for their full integration in a fully-functional and densely integrated quantum photonic circuit.}

\title{GaAs integrated quantum photonics: Towards compact and multi-functional quantum photonic integrated circuits}
\titlerunning{GaAs integrated quantum photonics}
\author{Christof P. Dietrich\inst{1,*}, Andrea Fiore\inst{2}, Mark G. Thompson\inst{3}, Martin Kamp\inst{1}, and Sven H\"{o}fling\inst{1,4}}
\authorrunning{C.P. Dietrich et al.}
\institute{
Technische Physik, Universit\"{a}t W\"{u}rzburg, Am Hubland, 97072 W\"{u}rzburg, Germany
\and
COBRA Research Institute, Eindhoven University of Technology, 5600 MB Eindhoven, The Netherlands
\and
Centre for Quantum Photonics, H. H. Wills Physics Laboratory and Department of Electrical and Electronic Engineering, University of Bristol, Woodland Road, Bristol BS8 1UB, United Kingdom
\and
SUPA, School of Physics and Astronomy, University of St Andrews, North Haugh, St Andrews KY16 9SS, United Kingdom
}
\mail{\email{christof.dietrich@physik.uni-wuerzburg.de}}

\maketitle

\section{Introduction}
Quantum information science using the unique features of quantum mechanics - superposition and entanglement - can greatly enhance computational efficiency, communication security, and measurement sensitivity. The generation of quantum bits (qubits) and the realization of quantum gates are the prerequisites for conducting quantum information science. The use of photons as qubits has been widely considered as one of the leading approaches to encode, transmit and process quantum information because of their low decoherence, high-speed transmission and compatibility with classical photonic technology\cite{OBrien2009,Faraon2011,Lodahl2015}. Photonic qubits can easily be encoded in many different degrees of freedom, e.g. polarization, path, frequency and orbital angular momentum, making the implementation of qubits using single photons very attractive. However, single photons are subject to substantial losses. Furthermore, nonlinear interactions between single photons are weak and do not provide $\pi$ cross phase modulation in natural nonlinear materials, which is the requirement for achieving a two-qubit gate.

In 2001, a major breakthrough known as the KLM scheme \cite{Knill2001} showed that scalable quantum computing is possible by only combining single-photon sources, linear optical circuits and single-photon detectors. In this case, the non-linearity in the detection replaces a direct optical non-linearity. The resulting quantum gates are probabilistic, but the probability of success can be made as close to one as desired by increasing the number of resources (ancilla photons). In addition, it was proposed that quantum logic networks can be simulated by simple one-way quantum computation using a particular class of entangled states, so called cluster states \cite{Raussendorf2003}. On a parallel line, a particularly simple class of linear-optics quantum simulators, boson sampling circuits, were theoretically shown to implement computational problems which are classically intractable \cite{Aaronson2010}. By following these approaches of only using linear optical components, two-qubit \cite{OBrien2003} and three-qubit gates \cite{Lanyon2009}, simple quantum algorithms \cite{Lanyon2007,Lu2007} as well as boson sampling \cite{Tillmann2013,Crespi2013,Spring2013,Broome2013} could already be experimentally demonstrated. Early demonstrations of linear-optics quantum processing have relied on inefficient and bulky single photon sources based on spontaneous parametric down conversion (SPDC), modest efficiency (at near-infrared wavelengths) single photon detectors based on avalanche photodiodes (APDs) or superconducting nanowires, and optical circuits with bulk optical elements. However, similar to integrated electronics, the practicality and scalability of quantum information technology ultimately requires the integration of individual components on a single chip. Integrated quantum photonics is emerging as a promising approach for future quantum information science, since it enables a substantial improvement in performance and complexity of quantum photonic circuits and provides routes to scalability by the on-chip generation, manipulation and detection of quantum states of light. Major progress has been made recently towards highly efficient integrated single photon sources, single photon detectors and integrated photonic circuits.

In particular, linear quantum photonic integrated circuits (QPICs) for two-photon interference, CNOT-gates and entanglement manipulation could already be demonstrated on various platforms and with various materials. These include e.g. silica-on-silicon \cite{Politi2008,Spring2013,Carolan2015}, laser direct-writing silica \cite{Marshall2009}, gallium nitride \cite{Xiong2011}, lithium niobate \cite{Jin2014}, silicon-on-insulator \cite{Bonneau2015,Harris2015} and GaAs \cite{Wang2014}. Indium arsenide/gallium arsenide (InAs/GaAs) quantum dots (QDs) are routinely embedded in photonic crystal waveguides/cavities and have been established as robust and efficient single-photon sources \cite{Englund2007b,Lund-Hansen2008,Schwagmann2011,Laucht2012,Hoang2012}. Moreover, high-efficiency superconducting nanowire single-photon detectors based on GaAs and Si waveguides have been successfully demonstrated \cite{Sprengers2011,Gerrits2011,Pernice2012,Reithmaier2013b}.

%Periodically-poled lithium niobate waveguides and silicon wire waveguides have furthermore been used to generate single-photon pairs via spontaneous parametric down-conversion \cite{Tanzilli2001} and spontaneous four-wave mixing \cite{Takesue2008}, respectively. 

Among all these platforms, GaAs is a well-known, mature material system for classical integrated photonics. It allows for the fabrication of low-loss waveguides and its high refractive index enables tight confinement of light and therefore compact devices and circuits. It has been employed in GHz modulators due to its high $\chi^{(2)}$ nonlinearity \cite{Walker1989}. In the context of quantum information science applications, the large electro-optic effect in GaAs makes it a very promising candidate for fast routing and manipulation of single-photons. Single-photon emission in GaAs can be realized by either spontaneous parametric down-conversion in waveguides \cite{Leo1999}, taking advantage of the high $\chi^{(2)}$, or by integrating single QDs into nanophotonic structures. Single-photon sources based on parametric generation of photon pairs and heralding must be operated at relatively low average photon numbers, resulting in a low single-photon probability, unless complex multiplexing schemes are employed \cite{Collins2013}. The QD approach, facilitated by the direct bandgap of GaAs, enables efficiencies close to 100\% \cite{Arcari2014} and represents a fundamental advantage of GaAs QPICs with respect to Si or LiNbO$_3$ circuits.  

Besides single-photon emission, also single-photon detectors on GaAs waveguides have already been successfully demonstrated\cite{Sprengers2011}. Therefore, GaAs waveguide circuits monolithically integrated with single-photon sources and superconducting single-photon detectors offer a promising approach to large-scale quantum photonic integrated circuits. Here, we review recent developments in InAs/GaAs QD single-photon sources (Sections 2-4), ridge waveguide quantum photonic circuits (Section 5) and GaAs waveguide single-photon detectors (Section 6), and we also discuss the challenge of monolithic integration of individual components (Section 7) towards the realization of dense and fully-functional quantum photonic integrated circuits. Recent promising developments in parametric generation of photon pairs \cite{Leo1999} and entangled photons \cite{Orieux2013}, including by electrical injection \cite{Boitier2014}, have been reviewed recently in Ref.\cite{Autebert2015} and are therefore not discussed here.

\begin{figure}
\center
\includegraphics[width=0.95\linewidth]{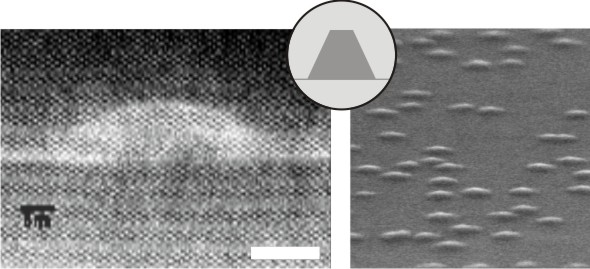}
\caption{\label{Fig1}Left: Transmission electron micrograph of an InAs QD (bright area) in a GaAs matrix (dark surrounding) grown by MBE in Stranski-Krastanov growth mode. The white scale bar represents 10\,nm. Right: SEM image of uncapped InGaAs QDs.}
\end{figure}

\section{Semiconductor quantum dots}
Integrated quantum photonic experiments rely on the generation, manipulation and detection of single photons that are ideally created on demand in a two-level system. Besides natural two-level systems such as single atoms, molecules or defect centers, also nanometer-sized inclusions of a low-gap semiconducting material incorporated into a high-gap matrix delivers a two-state configuration by forming a quasi zero-dimensional potential trap. Those energy traps are called semiconductor QDs (QDs). Single-photon sources based on QDs have the huge advantage that they can be grown monolithically with monolayer precision under controlled conditions by well-established growth techniques such as molecular beam epitaxy (MBE) \cite{Snyder1991,Leonard1993,Moison1994,Grundmann1995} or metal-organic vapor phase epitaxy (MOVPE) \cite{Oshinowo1994,Petroff1994,Heinrichsdorff1996}. The unique optical properties of semiconductor QDs have enabled extensive research and tremendously increased the scientific interest in realizing photonic QD devices. QDs are nowadays widely used in applications such as light-emitting diodes or solar cells. Regarding integrated quantum photonics, numerous proof-of-principle experiments were carried out on QDs, including single photon emission\cite{Michler2000}, two-photon interference \cite{Santori2002,Patel2010}, polarization-entanglement \cite{Akopian2006} and strong coupling \cite{Reithmaier2004,Yoshie2004}, underpinning their promising applicability as qubit source in quantum communication schemes.

The most extensively studied QD materials are InAs and InGaAs in GaAs matrices (see Figure \ref{Fig1}). The huge difference in band-gap energies between InAs ($E_\text{g}$(2\,K) = 0.422\,eV) and GaAs (1.522\,eV) and the three-dimensional quantum confinement makes it possible to tune the QD emission in a very large spectral window - from almost 850\,nm up to 1400\,nm - by adjusting the QD dimensions. In the following, we will highlight achievements in the growth of QDs (Section 2.1), their positioning on determined sites (Section 2.2) and their optical properties (Section 2.3).

\begin{figure}
\center
\includegraphics[width=0.75\linewidth]{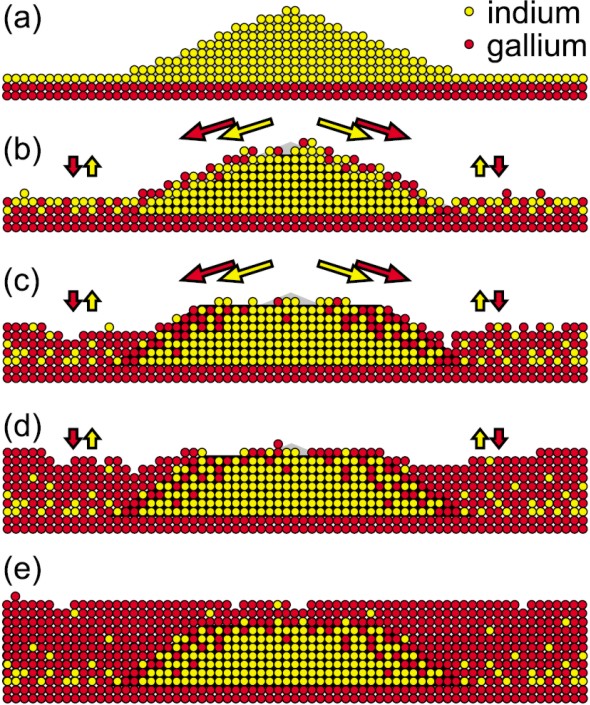}
\caption{\label{Fig2}Schematic capping process of a pyramidal InAs QD (a) overgrown by GaAs layers with increasing thicknesses (b-e). Reprinted with permission from \cite{Eisele2008}. \textcopyright 2008, AIP Publishing LLC.}
\end{figure}

\subsection{Self-assembled quantum dots}
The lattice mismatch between InAs and GaAs is about 7\% giving rise to considerable strain when the two materials are deposited on top of each other. This circumstance is used in the most exploited growth mode for QDs - the Stranski-Krastanov growth mode \cite{Stranski1938} - which makes use of the fact that coherent, dislocation-free islands form self-assembled as result of strain compensation. The size of the islands is thereby extremely sensitive to the amount of material that is deposited. 

The growth of InAs on a GaAs (100) surface initially results in a thin 2D wetting layer. Due to the lattice-mismatch, the two-dimensional growth mode turns into a three-dimensional growth after deposition of a few monolayers resulting in the creation of randomly positioned QDs with a pyramidal shape. In order to prevent the dots from oxidation and to separate them from the surface (for the integration into photonic nanostructures, see Section 4), they are commonly overgrown by a capping layer of GaAs completing the three-dimensional quantum confinement. The capping changes the shape of the QDs from purely pyramidal to truncated pyramidal (see Figure \ref{Fig2}) \cite{Eisele2008}. During the capping process, intermixing between the capping material and the QDs might occur leading to the formation of In(Ga)As QDs. This intermixing strongly affects the QD composition and potential profile as well as the QD emission.

In recent years, several techniques have been developed that facilitate the direct manipulation of QD properties including their size, their size distribution, their density, their shape and emission wavelength. The QD size can be varied by changing the composition of the dots\cite{Loeffler2006}. The size distribution of QDs directly determines the inhomogeneous linewidth of the QD ensembles: large size variations lead to an unwanted broadening of the emission. A way to reduce the size distribution spread would be partial capping and annealing: an almost uniform height of QDs can be achieved by introducing an annealing step during the process of QD capping \cite{Garcia1998}. This process is usually accompanied by a blueshift of the QD emission due to the height reduction. 

Besides the control of the uniformity, the growth conditions can be used to change the shape of the confinement potential and thereby the emission energy. Indeed, the emission properties of QDs are determined by the strength of the quantum confinement effect which can be altered in several different ways. Besides changing the vertical height of QDs by e.g. partial capping, the capping layer composition can be optimized to alter the strain state and heterostructure potential in the QD, or QDs can be annealed after growth leading to an inter-diffusion of gallium and indium at the QD surface. The diffusion process undermines the quantum confinement and causes a blueshift \cite{Garcia1998,Yang2007,Ellis2007} whereas an InGaAs capping layer reduces the strain in the QD and reduces the inter-diffusion, leading to a red-shift of the QD emission \cite{Nishi1999,Yeh2000}.

\begin{figure}
\center
\includegraphics[width=0.95\linewidth]{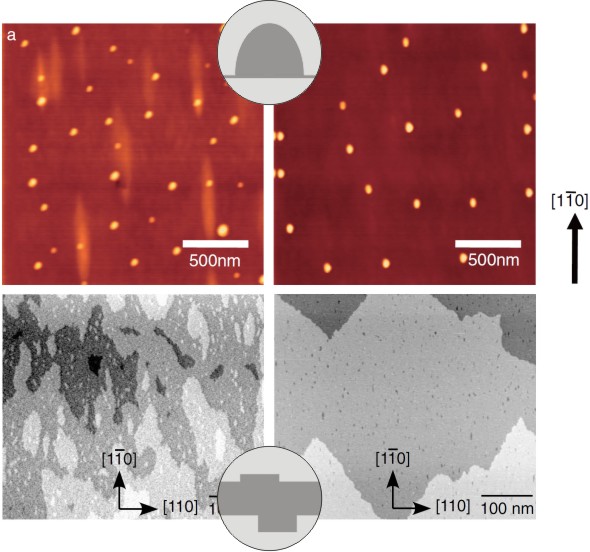}
\caption{\label{Fig3}Top row: Atomic force micrographs from QDs grown by droplet epitaxy before (left) and after annealing (right) at 400$^\circ$C. \textcopyright IOP Publishing. Reproduced with permission. All rights reserved. Bottom row: Scanning tunnel micrographs of monolayer fluctuations just after growth interruption for bottom and top QW interfaces (left and right). Adapted with permission from Ref.\cite{Peter2005}. Copyrighted by the American Physical Society.}
\end{figure}

The control of the density of QDs is crucial for the performance of single photon experiments since it requires the ability to optically address individual QDs by conventional spectroscopic methods. Several routes were already demonstrated including the change of growth parameters such as growth temperature \cite{Alloing2007}, III/V-ratio \cite{Li2013} or growth interruptions \cite{Convertino2004}. One of the most reliable ways to control the density of QDs (over several orders of magnitude) is the change of the growth rate, or by varying the amount of deposited material (approaching the 2D to 3D material growth transition range that can be controlled by systematically applying material flux gradients). In this way, densities of self-assembled QDs as low as $< 0.1\,\mu$m$^{-2}$ can be achieved \cite{Huang2007,Alloing2007}.

Alternatives to growing self-assembled QDs in the Stranski-Krastanov growth mode are techniques such as droplet epitaxy \cite{Koguchi1993} or the control of monolayer fluctuations in quantum wells \cite{Gammon1996,Hours2005} (see Figure \ref{Fig3}). Droplet epitaxy is based on the affinity of some group-III elements to form droplets on a semiconductor surface and has the advantage that it does not require a lattice-mismatch between the involved materials \cite{Mano2009,Tighineanu2013}. For droplet epitaxy, droplets of group-III elements are annealed after growth in an arsenic atmosphere and, under optimum annealing conditions, form defect-free QDs by saturating with As (see Figure \ref{Fig3} top). Compared to the Stranski-Krastanov growth, QDs grown by droplet epitaxy are relatively large. Alternatively, strain-free QDs can be achieved by interrupting the growth of thin quantum wells. These growth interruptions intentionally introduce monolayer fluctuations (Figure \ref{Fig3} bottom) at the interface effectively creating an additional quantum confinement within the quantum well plane. This type of QD has a larger lateral extension in the layer plane than QDs grown by the Stranski-Krastanov mode or droplet epitaxy \cite{Peter2005}, leading to a larger oscillator strength. They further benefit from the fact that they are not subject to material intermixing. 

\subsection{Site-controlled quantum dots}
The implementation of semiconductor QDs as sources of flying qubits into quantum integrated circuits requires a precise control over the relative position of the QD with respect to the circuit. Small misalignment (e.g. of a QD in the center of a defined PhC cavity) can lead to severe reductions in device efficiency.

\begin{figure}
\center
\includegraphics[width=0.75\linewidth]{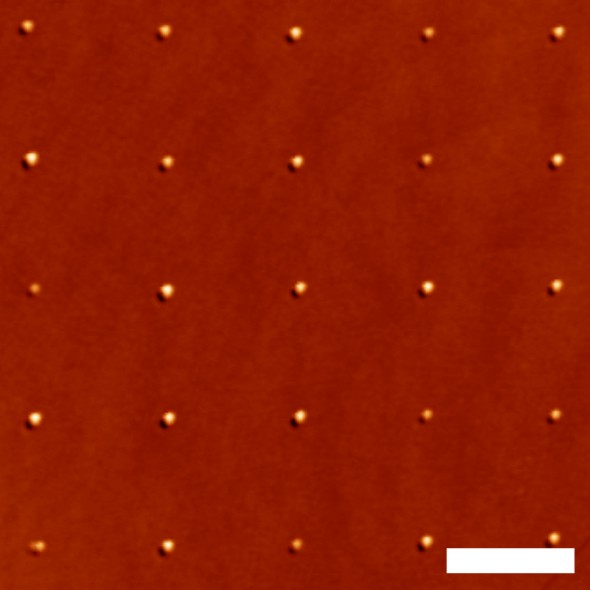}
\caption{\label{Fig4} Atomic force micrograph of site-controlled QDs with 1\,$\mu$m lattice period in a square lattice grown on a mesa structure. The scale bar represents 1\,$\mu$m. Reprinted with permission from \cite{Schneider2008}. \textcopyright 2008, AIP Publishing LLC.}
\end{figure}

Adequate control can be achieved by either placing the circuit around the random position of a self-assembled QD \cite{Badolato2005,Hennessy2007} or by predetermination of both the QD and circuit location. The first approach makes a pre-characterization of the QD distribution necessary, followed by a top-down etching around a chosen dot. However, this process is technologically very challenging and seems incompatible with the upscaling to large quantum photonic circuits, as the device layout must be adapted to the QD spatial positions. The growth of QDs on predetermined sites provides accurate alignment (see Figure \ref{Fig4}) of single photon sources with respect to photonic circuits \cite{Suenner2008} and facilitates high yield device fabrication. Several strategies were developed in the past to precisely position QDs on a semiconductor platform \cite{Schmidt2007}. This includes e.g. the deposition of optically active QDs on an optically inactive stress layer \cite{Krenner2005}. A more common strategy is the pre-patterning of the semiconductor surface by e.g. drilling holes into it \cite{Heidemeyer2004,Kiravittaya2006,Gallo2008,Mereni2009,Huggenberger2011a,Helfrich2012,Maier2014a,Unsleber2015b}. This can be achieved by a combination of electron beam lithography and ion etching, by atomic force nano-lithography \cite{MartinSanchez2009}, by nanoimprint lithography \cite{Tommila2012} or by local oxidation nano-lithography \cite{CanetFerrer2013}. During subsequent regrowth the adatoms preferentially nucleate at the patterned sites and form QDs. This process can provide an accuracy of the QD position of about $\pm50$\,nm which in many cases is sufficient for the application in photonic circuits \cite{Huggenberger2011b}. Similar results can be obtained by growing site-controlled QDs using masked surfaces \cite{Tatebayashi2000}.

In experiments and applications exploiting multiphoton interference (see section 3.3), QD emission linewidths are an important figure of merit and should ideally reach transform-limited values. In this regard, the linewidth has been found to strongly depend on the proximity to heterointerfaces and free surfaces, due to the effect of fluctuating charge states \cite{Houel2012}. For optimized growth conditions and advanced sample designs, record linewidth values as low as 7\,$\mu$eV were reported for randomly occupied QD sites \cite{Joens2013} and around 20\,$\mu$eV were demonstrated for QD patterns with only one QD at each site \cite{Mereni2009,Schneider2012} ($p$-shell excitation, see Section 3.1). In this regard, site-controlled QDs are comparable to self-assembled QDs in terms of emission linewidths. Another crucial feature for photonic applications is the spectral width of the QD ensemble, which is mainly determined by size and shape fluctuations of the grown dots. Whereas self-assembled QDs exhibit inhomogeneous broadening in the range of several tens of meV, the nucleation on predetermined sites can reduce this value to only a few meV \cite{Mohan2010}.
 
\subsection{Optical properties of quantum dots}
Most common barrier and QD material combinations, including the extensively investigated combination of InAs QDs in GaAs barriers, form a type-I heterostructure resulting in strong quantum confinement for both electrons and holes. In good approximation, it can be assumed that the QD behaves as a two-level system, where the lowest-energy electronic excitation (heavy-hole exciton, X) involves one electron in the conduction band and one hole in the valence band with strong heavy-hole character (note that by applying elastic stress to initially unstrained QDs the excitonic ground state can also have light-hole character \cite{Huo2014}). Higher occupied states such as biexcitons or trions are detuned in energy due to Coulomb interactions. The degeneracy of the QD valence band is usually lifted by the asymmetric shape of the confinement potential and by strain causing a non-vanishing energy splitting between heavy and light hole levels on the order of several tens of meV. For typical self-assembled QDs the splitting is so large that the transition between conduction band and light-hole valence band can easily be neglected making the system a true two-level system.

\begin{figure}
\center
\includegraphics[width=0.95\linewidth]{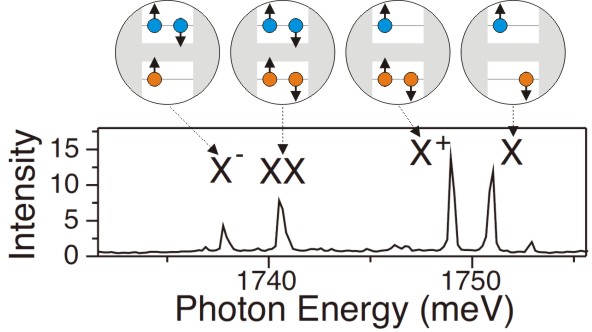}
\caption{\label{Fig5} Non-resonant photoluminescence spectrum of a GaAs QD showing exciton (X), biexciton (XX) and trion (X$^{-}$, X$^{+}$) transitions. The upper panel schematically depicts the QD charge configuration for each transition. Adapted with permission from Ref.\cite{Abbarchi2010}. Copyrighted by the American Physical Society.}
\end{figure}

The total angular momentum $J$ of heavy-hole excitons is a sum of heavy-hole angular momentum and electron spin allowing four different transition configurations having an angular momentum along the growth direction: $J_{z,1} = \pm1$ and $J_{z,2} = \pm2$. The $J_{z,1}$ configurations are termed bright exciton states as they couple to the optical field and therefore preferentially decay through radiative recombination channels. In contrast, configurations with total angular momentum $J_{z,2}$ are termed dark exciton states and decay through non-radiative recombination channels. Due to the close proximity of both electrons and holes in semiconductor QDs, exchange interactions of electron-hole pairs are enhanced and cause an energetic splitting between dark and bright states as well as the formation of mixed dark and mixed bright states. Bright and dark exciton decay bi-exponentially caused by radiative and non-radiative recombination as well as spin flip processes (turning bright states into dark states, and vice versa, under creation or annihilation of LA phonons) \cite{Roszak2007,Johansen2010}.

While the pure states are circularly polarized, they mix as a consequence of exchange interactions and of the preferential QD elongation along the in-plane $[1\overline{1}0]$-direction for growth on GaAs(100) substrates, giving rise to mixed states with linear polarization planes parallel to the $[1\overline{1}0]$- and $[110]$-crystal direction \cite{Seguin2005,Bennett2010}. Those linearly polarized transitions can directly be recorded in luminescence experiments \cite{Bayer2002}. Their energetic separation is the so-called fine structure splitting (FSS). Dark states have typical fine structure splittings in the order of 1$\mu$eV \cite{Poem2010} whereas bright states can have FSS up to several hundreds of $\mu$eV\cite{Seguin2005} (few to few tens of $\mu$eV are more commonly observed).

Besides bright and dark exciton states, also states with higher occupation such as trions (i.e. charged excitons, X$^{-}$ or X$^{+}$) or biexcitons (XX) occur. Owing to Coulomb interactions between confined carriers, these states have recombination energies different from the exciton transitions allowing the spectral filtering of individual excitonic transitions. A typical QD spectrum showing the radiative transitions of excitons (X), biexcitons (XX) as well as trions (X$^{-}$, X$^{+}$) can be seen in Fig.\ref{Fig5}.

Biexcitons are particularly interesting since they posses a net projection of the angular momentum of 0 and are therefore not subject to exchange interactions. If the fine-structure splitting is suppressed, this leads to the decay into two bright excitons with opposite circular polarization. This decay scheme has extensively been investigated for the creation of polarization-entangled photon pairs in QDs systems\cite{Benson2000,Stevenson2006,Akopian2006,Juska2013}. For the entanglement, the two decay paths $\sigma^{+}_\text{XX}\rightarrow\sigma^{-}_\text{X}$ and $\sigma^{-}_\text{XX}\rightarrow\sigma^{+}_\text{X}$ (with $\sigma$ being the photon polarization state) need to be indistinguishable. This is typically not the case due to finite fine structure splittings. Therefore, in QD systems with fine structure splittings smaller than the exciton linewidths (as for QDs grown on GaAs(111) layers \cite{Kuroda2013}), entangled photon pairs can deterministically be created \cite{Larque2008}. Other techniques such as thermal annealing \cite{Young2005}, the application of electric fields \cite{Kowalik2005}, strain \cite{Trotta2012}, as well as the shaping of the confinement potential \cite{Juska2013} are also able to reduce the FSS and produce entangled photon pairs.

%Note that biexcitons decay single exponentially owing to the absence of dark states. Biexcitons also decay faster than single exciton states since two decay channels are available. A typical quantum dot spectrum can be seen in Figure \ref{Fig5} showing the radiative transitions of excitons (X), biexcitons (XX) as well as trions (X$^{-}$, X$^{+}$). 

\section{Quantum dots as single photon sources}
On-chip single-photon sources can be realized by exploiting the radiative recombination from an excitonic state of a single QD \cite{Michler2000}. Single QDs have been widely investigated as single-photon and entangled-photon sources (see \cite{Shields2007} for a review). As compared to single-photon sources based on parametric down-conversion and heralding, QD-based sources present the advantage of deterministic, on-demand single photon emission, much easier filtering of the pump due to the lower powers needed (tens of nW level) and the possibility of pumping from the top - this is particularly relevant in the context of quantum photonic integrated circuits as it reduces the coupling of pump light into the chip.

\subsection{First- and second-order coherence}
In general, quantum emitters can be classified regarding their coherence properties and photon statistics. The degree of first-order coherence is quantified by the field-field correlation function $g^{(1)}$ and the degree of second-order coherence is described by the intensity-intensity correlation function $g^{(2)}$. $g^{(1)}$ of a quantum emitter is usually probed in optical interferometers where the optical beam is split into two arms and joined again after introducing a variable time delay $\tau$ into one of the arms. The light produced by an ideal single-quantum emitter with a lifetime $T_1$ is described by $\left|g^{(1)}(\tau)\right|$ which decays exponentially with a time constant $T_2$ = $2T_1$ - corresponding to a single-photon pulse with no phase jumps. However, real quantum emitters such as excitons in QDs are imperfect systems and show faster decays of $\left|g^{(1)}(\tau)\right|$ caused by dephasing mechanisms such as phonon interactions. Dephasing reduces the coherence time of the QD resulting in $1/T_2 = 1/2T_1 + 1/T_2^*$ (with $T_2^*$ being the pure dephasing time) and increases its linewidth. Dephasing processes are detrimental in quantum photonic applications as they make photons emitted from different sources distinguishable and reduce the visibility in two-photon interference experiments.

\begin{figure}
\center
\includegraphics[width=0.95\linewidth]{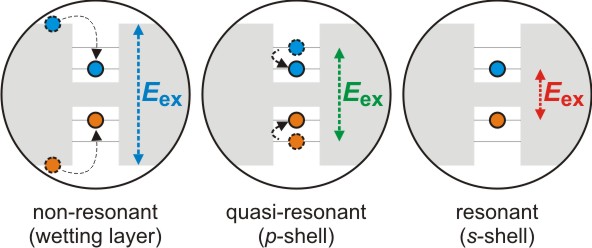}
\caption{\label{Fig6} Non-resonant, quasi-resonant and resonant excitation scheme: high-energy particles (dotted circles) decay through phonon-interactions into the QD ground state (solid circles). The vertical arrows indicate the energy necessary for each process.}
\end{figure}

\begin{figure}
\center
\includegraphics[width=0.95\linewidth]{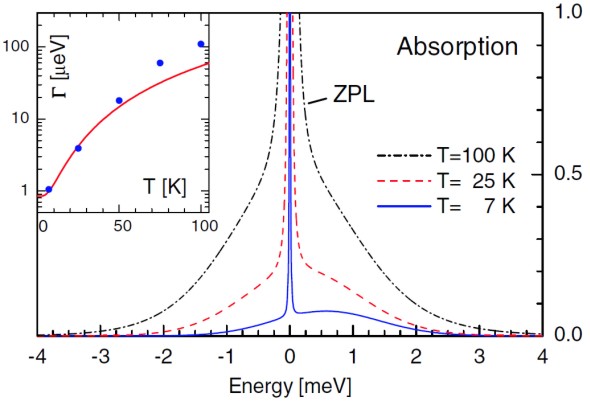}
\caption{\label{Fig7} Calculated absorption spectra of an InAs QD at different temperatures. At low temperatures, a clear asymmetric broadening of the zero-phonon line (ZPL) caused by exciton-phonon-interactions with longitudinal acoustic phonons is visible. Inset: Calculated broadening of the ZPL compared with experimental results from \cite{Borri2001}. Reprinted with permission from Ref.\cite{Muljarov2004}. Copyrighted by the American Physical Society.}
\end{figure}

\begin{figure}
\center
\includegraphics[width=0.95\linewidth]{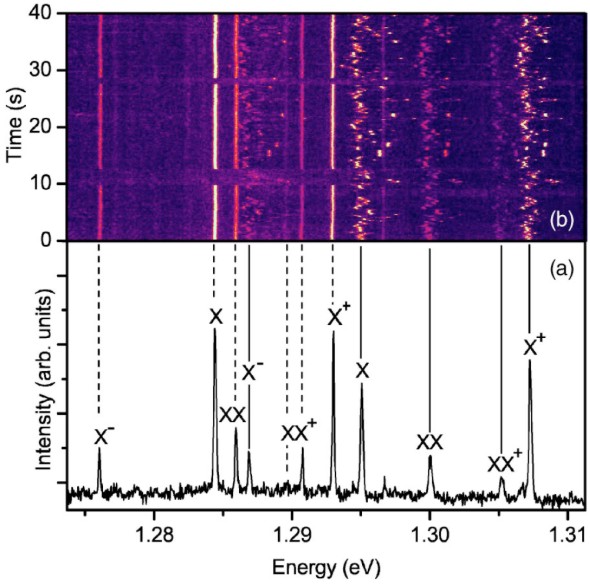}
\caption{\label{Fig8} (a) Typical spectrum of InAs/GaAs QDs measured through an aperture with 100\,nm width. (b) Temporal evolution of (a) showing spectral wandering of almost all QD transitions. Reprinted with permission from Ref.\cite{Rodt2005}. Copyrighted by the American Physical Society.}
\end{figure}

The occurrence of dephasing is strongly related to the excitation scheme used for the experiment. Nowadays, three main different schemes are distinguished: off-resonant, quasi-resonant ($p$-shell excitation) and resonant pumping ($s$-shell excitation), see Figure \ref{Fig6} for the different energy configurations (less common excitation schemes omitted here include wetting layer excitation or excitation above shell resonant to an LO phonon replica). The excitation of a QD well above its ground state or resonant with an excited state (e.g. $p$-shell) results in the relaxation of carriers into the ground state by generation of phonons \cite{Borri2001}. In general, QDs always emit and absorb phonons through inelastic processes. Especially, LA phonons are well known sources for dephasing and cause an asymmetric broadening of the QD transition (as visible in the calculated absorption spectrum shown in Figure \ref{Fig7}) \cite{Muljarov2004,Ramsay2010b,Madsen2013,Kaer2013,Kaer2014}. The interaction of QD excitons with phonons can either be suppressed by performing the experiment at very low temperatures or by adapting resonant $s$-shell excitation (see Section 3.2). We note that off-resonant and quasi-resonant excitation also produce a certain time jitter in the generation of single photon events, reducing the indistinguishability \cite{Ates2009,Flagg2012,Unsleber2015}. Another dephasing process is induced by the random distribution of charge carriers inside a QD and changes in the electronic configuration around the QD. Charge and spin fluctuations result in a persistent change of the excitonic ground state inducing spectral wandering of the QD emission (see Figure \ref{Fig8}). This leads to an effective increase of the QD linewidth and a decrease of the QD coherence time \cite{Kuhlmann2013}. Charge fluctuations can be suppressed drastically by employing resonant pumping and adding a weak auxiliary continuous wave reference beam to the excitation beam of the QD\cite{Konthasinghe2012b}. Spectral wandering is typically much slower than the radiative decay of excitons and its effect can in some cases be circumvented by pumping the QD with a short pulse twice at short time intervals to obtain two photons with very similar energies \cite{He2013b}. However, when interference between photons from different QDs is required, spectral wandering produces a loss of indistinguishability in the same way as pure dephasing, and must be suppressed \cite{Gao2013,Gold2014}.


Photon statistics can be probed by measuring the second-order correlation function $g^{(2)}(\tau)$ that gives the probability of detecting two photon events with time delay $\tau$ between them. In this regard, the $g^{(2)}(0)$ value (at zero time delay) is particularly important since it measures the probability that two photons are detected at the same time. Experimentally, $g^{(2)}(\tau)$ is routinely determined in a Hanbury Brown and Twiss (HBT) setup that consists of a 50:50 beam splitter and two nominally equal single-photon detectors with a temporal resolution much better than $\tau$. Whereas for a coherent and thermal light source $g^{(2)}(0)$ approaches 1 and 2, respectively, a perfect single photon source gives $g^{(2)}(0) = 0$ since the probability of detecting two photons events at the same time is then zero.

The photon statistics of real single-photon sources are strongly affected by non-idealities such as multi-photon emission from multi-exciton states, collection of the emission of other QDs, re-pumping of the emitter after the emission of the first photon, etc., and therefore produce non-zero coincidences\cite{Kaer2013,Kaer2014}. Resonant, pulsed excitation can lead to $g^{(2)}(0)$ values very close to zero \cite{He2013b,Ding2016,Somaschi2016}. Pumping by adiabatic rapid passage using chirped laser pulses can further suppress multi-photon emission \cite{Wei2014}. Even with these highly sophisticated excitation and detection schemes, multi-photon emission in QD systems can never be turned off completely. In this regard, the measured $g^{(2)}(0)$ value can be seen as a measure of the quality of the single-photon source \cite{Ding2016,Somaschi2016}.

\subsection{Resonance fluorescence}
As mentioned above, resonant $s$-shell excitation circumvents several dephasing processes, such as the creation of high-energy carriers or the interaction with phonons, leading to a considerable increase of the coherence time of QDs \cite{Birkedahl2001}. However, the monochromatic pumping of a two-level system at resonance offers many more advantages including the coherent generation \cite{Muller2007,Ates2009,Flagg2009,Vamivakas2009,Moelbjerg2012}, manipulation \cite{Vamivakas2009,Ulrich2011,He2013a,Peiris2014} and characterization \cite{Lu2010} of the excitonic states. 

\begin{figure}
\center
\includegraphics[width=0.95\linewidth]{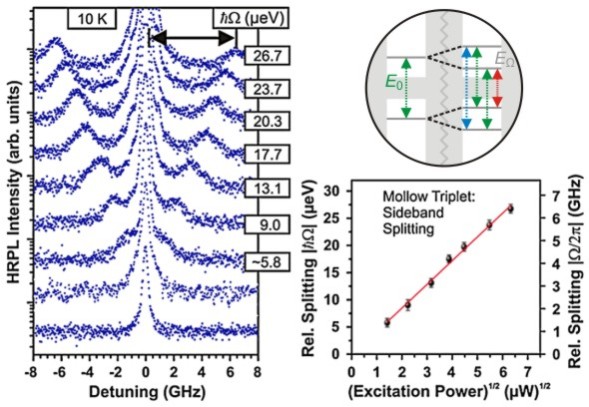}
\caption{\label{Fig9} Left: Excitation-dependent PL spectra of a resonantly-pumped single QD. $E_\Omega =\hbar\Omega$ denotes the energetic splitting between zero-phonon line and its first Mollow side band. Top right: Evolution of QD transitions in the "dressed" picture. Bottom right: The determined sideband splitting $\hbar\Omega$ vs. square root of the excitation power showing a linear dependence. Adapted with permission from Ref.\cite{Ates2009}. Copyrighted by the American Physical Society.}
\end{figure}

In general, the spectral and temporal properties of photons scattered by a two-level system are determined by the laser detuning from the excitonic resonance. The electric-dipole interaction is responsible for a coupling of the excitonic states to the driving field and turns the two levels of the exciton into "dressed states" \cite{Xu2007}. The dressed-state picture allows four radiative transitions of which two are degenerate and two are spectrally displaced by the Rabi energy $E_\Omega$ (see Figure \ref{Fig9} top right). This results in the emergence of three emission lines, the so-called Mollow triplet \cite{Mollow1969}, with the central line having the highest intensity (Figure \ref{Fig9} left). The Rabi energy scales with the excitation density of the driving field (as shown in Figure \ref{Fig9} bottom right). This enables a direct control of the side band transitions. By introducing a second laser beam in resonance with one of the sideband transitions (preferably the lower energy sideband), the Mollow triplet turns into a multitude of peaks under suppression of the main peak (due to deconstructive interference). In these multiply dressed QD states also phenomena like the multi-photon AC Stark effect can be present that have previously only been observed in atomic physics \cite{He2015}.

Whereas the above mentioned phenomena can be observed by using continuous laser irradiation, pulsed resonant driving of a two-level system (with ultra-short pulses) further facilitates the external control of the QD population as a consequence of the oscillatory behavior of the QD population versus excitation pulse area (so-called Rabi oscillations \cite{Kamada2001,Stievater2001}). In this way, the creation of single photons can drastically be increased by applying pulse areas matching the maximum QD population ($\pi$-pulses). 

In the Mollow triplet regime, most of the light is scattered incoherently owing to the saturation of the QDs. Reducing the strength of the driving field changes the scattering properties of QDs \cite{Konthasinghe2012b}. In the so-called Heitler regime (at very low excitation powers) the QD behaves like a passive scatterer and incident photons are scattered almost fully coherently \cite{Mollow1969}. As a consequence, the spectrum of coherently scattered photons is only determined by the spectral characteristics of the excitation source \cite{Matthiesen2013}. In this way, QDs linewidths as low as 0.03\,$\mu$eV have been observed already. Those QDs are in the "subnatural linewidth" \cite{Matthiesen2012} or "ultracoherence" regime \cite{Nguyen2011}. The application of pulsed excitation schemes in combination with the weak coherent scattering regimes enables pure QD spectroscopy in the absence of almost all dephasing mechanism and, with that, facilitates the deterministic generation of single photons. However, the low photon generation rates are a disadvantage. At moderate excitation powers, coherent scattering and the Mollow triplet can also coexist when the resonantly driven dot is thermalized with the phonon bath \cite{McCutcheon2013}.

Experimentally, the observation of resonance fluorescence is challenging. Sophisticated resonant electrical injection is possible \cite{Conterio2013} but technologically very challenging. More common optical pumping schemes suffer from problems related to the separation of the excitonic emission from the scattered laser light requiring elaborated sample designs or advanced excitation and detection schemes. A simple way of performing optically pumped resonance fluorescence experiments is the use of planar waveguides \cite{Muller2007,Melet2008,Jayakumar2013}. The laser light is coupled to the waveguide mode exciting the QDs which emit perpendicular to the sample plane. Another approach to the pump suppression is the cross-polarization technique that requires a high extinction rate between both polarizations \cite{Vamivakas2009}. In general, the detection of resonance fluorescence is always aggravated by scattering at imperfections. This circumstance becomes even more delicate in photonic nanostructures such as ridge waveguides\cite{Makhonin2014,Reithmaier2015} and photonic crystals \cite{Faraon2008b,Reinhard2012}. (Pulsed) resonance fluorescence has so far not been observed in waveguide-coupled photonic crystal cavities.

\begin{figure}
\center
\includegraphics[width=0.95\linewidth]{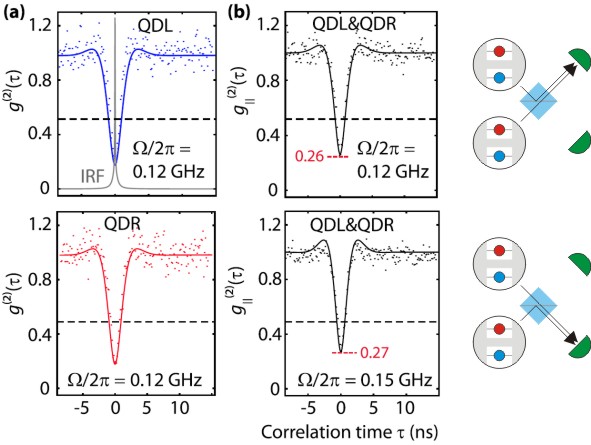}
\caption{\label{Fig10} (a) Second-order correlation function $g^{(2)}(\tau)$ for two QDs on the same chip separated by 40\,$\mu$m. (b) Two-photon interference of the two QDs from (a). The grey line (IRF) in (a) displays the instrument response function. Adapted with permission from Ref.\cite{Konthasinghe2012a}. Copyrighted by the American Physical Society.}
\end{figure}

\subsection{Two-photon interference}
When two single photons with the same Fourier-transform limited spectrum, temporal profile and polarization from two spatially separate sources propagate with a linear network, one cannot distinguish which of the two photons stems from one or the other source, i.e. they are indistinguishable \cite{Bylander2003}. Due to the interference of the probability amplitudes, two indistinguishable photons impinging on the two inputs of a balanced beam splitter (50:50) will always exit together (so-called photon-bunching). The experimental demonstration of photon-bunching was first achieved by Hong, Ou and Mandel in 1987 \cite{Hong1987} and is one of the main prerequisites for quantum information science being the basis for e.g. KLM quantum gates\cite{Politi2008} and boson sampling\cite{Aaronson2010}. This so-called two-photon interference is usually measured using a Michelson interferometer consisting of a beam splitter, two distinct optical paths and two single-photon detectors at the two outputs of the splitter (an equivalent configuration based on a Mach-Zehnder interferometer can also be used). A temporal displacement between both inputs can be introduced either by changing the position of the beam splitter \cite{Hong1987} or by introducing a variable delay line into one of the photon paths \cite{Patel2008} for making the photons gradually distinguishable. For single photons from semiconductor QDs, two-photon interference was first demonstrated for photons from the same source by exciting the QD with two short pulses temporally separated by a 2\,ns delay\cite{Santori2002}. A variable mirror able to compensate for the delay time was placed at one of the beam splitter outputs, a fixed mirror at another output. In this way, five different scenarios were created of photons impinging on the detectors of which only one creates a temporal overlap of both photons showing a dip in the correlation function. So far, two-photon interference could be demonstrated for very different systems including dissimilar (between a QD and a Poissonian laser \cite{Bennett2009}, a parametric down-conversion source \cite{Polyakov2011} or a frequency comb \cite{Konthasinghe2014}) and similar single photon sources (between different QDs) \cite{Ates2012,He2013a}. In most cases, indistinguishability has been demonstrated between QDs of two different and spatially separate samples \cite{Patel2010,Gao2013,Gold2014}, but in the perspective of quantum photonic integrated circuits it is of special importance to demonstrate two-photon interference between single photon sources on the same semiconductor chip (Figure \ref{Fig10}). This has been achieved for two QDs on the same substrate with only 40\,$\mu$m distance between them \cite{Konthasinghe2012a}. The experimental difficulty here is to find two QDs within the field of view that have almost identical emission energies, linewidths and polarization. A much more practical approach is the tuning of the excitonic QD transitions by changing the properties externally, e.g. by applying an electric field. See section 4.3 for a review about tuning mechanisms.

\section{Purcell-enhanced single-photon emission}
Intrinsic dephasing processes (e.g. phonon scattering) and extrinsic spectral fluctuations, related to varying charge environment in the vicinity of the QD, typically limit the coherence time of single QDs to $T_2 \approx 100-200$\,ps, while the natural exciton lifetimes are limited to $T_1 > 400$\,ps. The exploitation of two-photon interference in photonic quantum information processing (QIP) requires the reduction of the lifetimes in order to achieve $T_2 \approx 2T_1$. This can be achieved employing cavity quantum electrodynamic effects in nanophotonic structures that allow the precise tailoring of the electromagnetic field surrounding of the dot. In this way, the rate of radiative recombination from the embedded emitter can be controlled via the Purcell effect \cite{Purcell1946} in e.g. a cavity. The enhancement factor of the spontaneous emission rate is hereby determined by the spatial and spectral matching between emitter and cavity mode, the quality factor $Q$ and the mode volume $V$ of the cavity mode in the photonic nanostructure  \cite{Kleppner1981}. In particular, for a spectrally narrow source in resonance with a broader optical mode, the Purcell-enhancement is proportional to the $Q/V$ ratio (bad cavity regime). In order to approach an almost perfect cavity-emitter interface by increasing the light-matter interaction, photonic nanostructures need to provide very high quality factors and low mode volumes.

\begin{figure*}
\center
\includegraphics[width=0.67\linewidth]{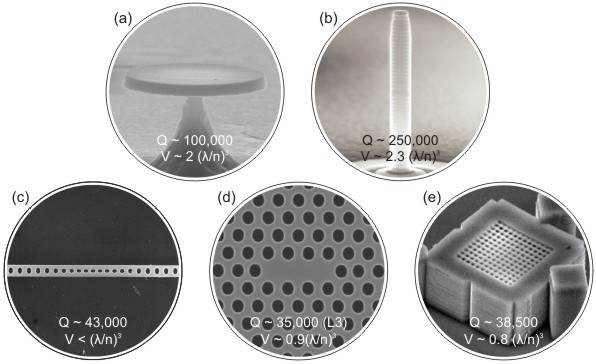}
\caption{\label{Fig11} Overview over different GaAs-based cavity types providing three-dimensional photonic confinement: (a) microdisks \cite{Balram2014}, (b) micropillars, (c) nanobeams \cite{Balram2014}, (d) PhC slabs and (e) three-dimensional photonic crystals. The respective record optical $Q$-factors for GaAs-based photonic nanocavities embedding QDs are given at the bottom of each image together with estimated mode volumes. (b) Adapted with permission from Ref.\cite{Reithmaier2004}. Copyrighted by the American Physical Society. (d,e) Adapted by permission from Macmillan Publishers Ltd: Sci. Rep. \cite{Luxmoore2013c}, \textcopyright 2013, and Nat. Photon. \cite{Tandaechanurat2011}, \textcopyright 2011.}
\end{figure*}

\subsection{Cavities}
Typical examples of photonic nanostructures include microdisks, micropillars, nanobeams, photonic crystal slabs and 3D photonic crystals (see Figure \ref{Fig11} for an overview). All structures provide an effective three-dimensional photonic confinement based on the possibility to manipulate and control light propagation by introducing interfaces with very high refractive index contrast. Obviously, the best index contrast for GaAs-based devices is given for the interface between GaAs (with $n = 3.5$) and air ($n = 1$). 

In microdisks, light is confined by total internal reflection (TIR) at the GaAs/air boundary and propagates within the disk cross section in the vicinity of the disk rim. The so-called whispering-gallery modes possess very high quality factors of up to $Q = 100,000$ \cite{Srinivasan2007} but suffer from the relatively large mode volumes. Even higher $Q$-factors can be observed in micropillar resonators ($Q$ can be as high as 250,000 \cite{Schneider2015}) where Bragg mirrors (alternating pairs of dielectric materials with high index contrast and $\lambda/4$-thickness for constructive interference) below and on top of the cavity region confine photons in the pillar effectively. Depending on the pillar diameters mode volumes as low as $2.3(\lambda/n)^3$ \cite{Lermer2012} can be achieved. However, while micropillars with embedded QDs are ideal vertical-emitting single-photon sources, they do not match the requirements of quantum photonic integrated circuits that involve in-plane photon routing and processing. In this regard, photonic crystals slabs embody the perfect compromise by facilitating high quality factors, low mode volumes and in-plane control. They are based on Bragg scattering in periodic photonic structures which are most commonly realized by drilling air holes in a lattice-geometry into a membrane of semiconductor material. In-plane confinement is then achieved by the photonic band gap introduced by the holes etched into the membrane, while out-of-plane confinement fully relies on total internal reflection at the membrane-air interface. Photonic crystals can be fabricated in square or triangular lattice whereas the latter is most commonly preferred as it provides a larger photonic band gap\cite{Joannopoulos2008}. 

PhC cavities are formed by displacing neighboring holes (H0 cavity \cite{Nozaki2010}) or leaving out a single (H1 cavity \cite{Englund2005,Shirane2007}) or several air holes, thereby forming a defect in the photonic crystal bandgap. In the prospect of single-photon emission, L3 cavities (three missing holes in a line) are most promising as they support polarized single modes with a wide spectral margin \cite{Fan2010}. The highest $Q$-factors for GaAs were so far reported for passive L3 cavities in a triangular lattice with reduced hole radius \cite{Saucer2013} and shifted next neighbor positions \cite{Akahane2003} with experimental values up to 700,000 \cite{Akahane2005,Combrie2008} (calculated Q-values exceed $1\times10^6$ for optimised L3 designs \cite{Minkov2014}). Active structures (with embedded QDs) show quality factors up to $Q$ = 35,000 (25,000) and mode volumes $V/(\lambda/n)^3 = 0.9 (0.4)$ for L3 (H1) cavities \cite{Takagi2012}. Note that the difference between $Q$-values with and without QDs may be related to the different wavelength range of operation (typically around 900\,nm with QDs, therefore in the spectral region of band-tail absorption in GaAs) or to residual non-resonant absorption by the QDs. Comparable results have been obtained using one-dimensional PhCs in nanobeams \cite{Enderlin2012,Balram2014}.

Three-dimensional PhCs, for example based on the woodpile structure \cite{Tandaechanurat2011} (see Figure \ref{Fig11}), may, in principle, offer higher quality factors due to suppressed leakage in the vertical direction and additionally provide better control of the QD spontaneous emission. However, they are technologically much more challenging to fabricate.

\begin{figure}
\center
\includegraphics[width=0.99\linewidth]{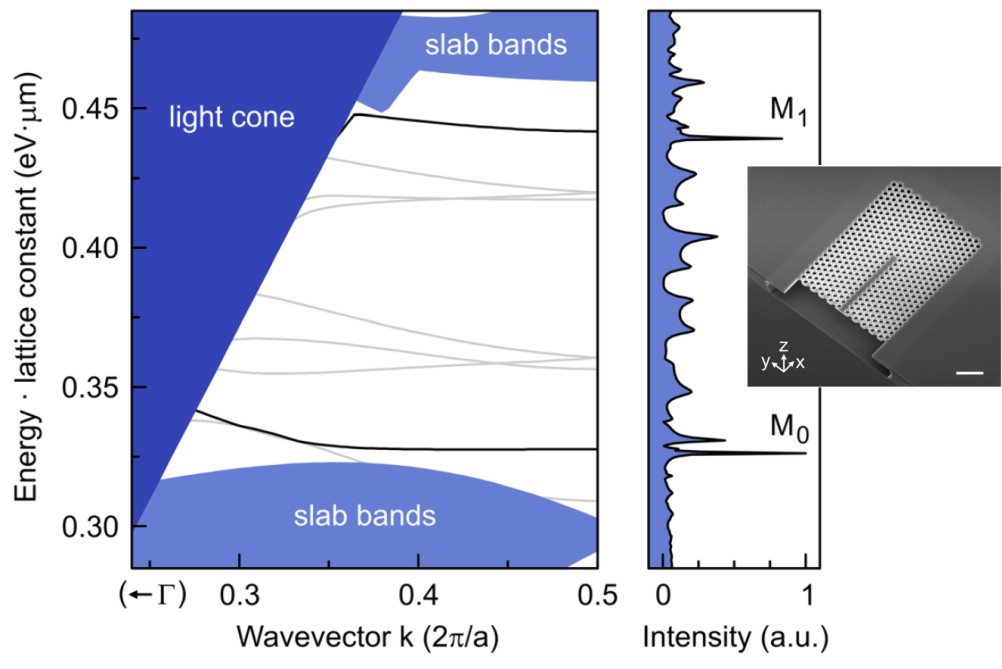}
\caption{\label{Fig12} Left: Calculated band structure (inside first Brillouin zone) for TE-modes of an infinite PhC waveguide. The black lines indicate $y$-polarized modes, the gray lines indicate $x$-polarized modes. Right: Scanning electron microscope image of the PhC waveguide. The scale bar represents 1\,$\mu$m. Reprinted with permission from \cite{Schwagmann2011}. \textcopyright 2011, AIP Publishing LLC.}
\end{figure}

\subsection*{4.2 Photon collection and routing in waveguides}
Quantum photonic integrated circuits are two-dimensional arrangements of different functionalities on one semiconductor chip that allow the generation, manipulation and detection of single photons. This requires efficient photon collection from single photon sources, interconnections between the individual functionalities as well as low-loss transport of light within the circuit from the stationary qubit to the manipulation and detection sites. Waveguides (such as PhC waveguides \cite{ViasnoffSchwoob2005,Englund2007b,Schwagmann2011,Laucht2012,Hoang2012} or ridge waveguides, see Section 5) perfectly meet these requirements as they efficiently collect photons emitted by the QDs (even in the absence of a cavity) and transport them via propagating modes. PhC waveguides have the huge advantage of a tight mode confinement paving the way for a high density of components on one chip. On the downside, they suffer from optical losses and mechanical instabilities. 

One way of efficiently collecting photons in a single guided mode is the use of a PhC waveguide \cite{Lecamp2007,MangaRao2007a,MangaRao2007b} (e.g. "W1" waveguide, based on a missing row of holes in a triangular lattice of holes patterned in a semiconductor slab). The radiative bandgap created by the PhC suppresses the emission into in-plane modes other than the intended mode (see Figure \ref{Fig12}), resulting in a large ($\beta = 80-90$\%) spontaneous emission coupling factor (defined as the fraction of spontaneous emission coupled to the desired mode). This is true even in the absence of a strong Purcell effect. The $\beta$-factor can be even higher in case the QD is spectrally close to the bottom of the dispersion curve. Then the emission into the guided mode is enhanced due to a decreased group velocity, further increasing $\beta$. Efficient funneling of the emission of single QDs into a PhC waveguide mode has been observed experimentally for QDs emitting in the 900\,nm \cite{Laucht2012} range but also for the telecommunication range around 1300\,nm \cite{Hoang2012}. In this context, $\beta$-factors as high as 0.98 \cite{Lund-Hansen2008,Arcari2014} and Purcell enhancements of up to $F_\text{p}$ = 2.7 \cite{Dewhurst2010} were already reported. %However, Arcari {\it et al.} pointed out that previously used top detection schemes as well as temperature tuning mechanisms may lead to a false measurement of the uncoupled QD lifetime in the waveguide and with that to an overestimation of the $\beta$-factor. By accounting for this effect, they were able to achieve near-unity $\beta$-factors (98.4\%) \cite{Arcari2014}.

\subsection{Tuning Quantum Dot and Cavity Mode}
\begin{figure}
\center
\includegraphics[width=0.95\linewidth]{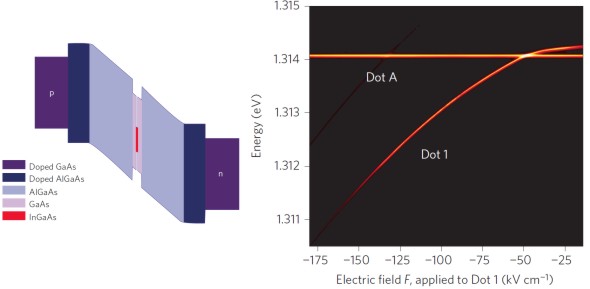}
\caption{\label{Fig13} Left: Band structure of a Stark-tunable device consisting of InAs QDs grown within GaAs quantum wells and AlGaAs superlattices. Right: Field-dependence of two QDs (Dot1 and DotA) as well as the cavity mode. Reprinted with permission from Macmillan Publishers Ltd: Nat. Photon. \cite{Patel2010}, \textcopyright 2010.}
\end{figure}

In a perfect cavity-emitter interface, both QD transition and cavity mode have the same energy for maximum Purcell enhancement. However, in real photonic nanostructures with embedded QDs it is most likely that both resonances are slightly detuned from each other as result of growth and processing imperfections. This circumstance makes external tuning mechanisms indispensable. The simplest way of tuning the QD excitonic emission with respect to the cavity mode is by changing the sample temperature in a temperature-variable cryostat \cite{Yoshie2004,Geveaux2006,Englund2007b}, by laser irradiation or by implementing heat pads close to the photonic crystal \cite{Faraon2007,Faraon2009}. All these mechanisms will affect the electronic band structure, the refractive index of the device and the lateral expansion of the cavity. The two latter changes only slightly affect the spectral position of the cavity mode, while the bandgap change strongly tunes the QD energy. Experimentally more challenging are other post-growth tuning mechanisms such as the application of external magnetic fields\cite{Kim2011} or strain \cite{Beetz2013,Sun2013}. A more integrated approach is the implementation of electrical contacts into the photonic nanostructures and shifting the QD transitions via Stark tuning by applying an electric field across the QD layer \cite{Fry2000,Hofbauer2007,Chauvin2009,Laucht2009,Patel2010,Hoang2012,Thon2011,Pagliano2014}. Additional heterostructure barriers (see Figure \ref{Fig13}) around the QD suppress carrier tunneling out of the dot enabling a tuning range of several tens of meV \cite{Patel2010}.

\begin{figure}
\center
\includegraphics[width=0.95\linewidth]{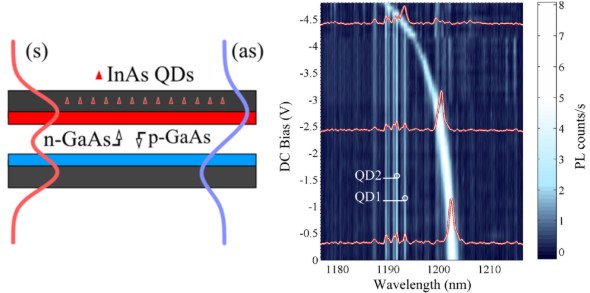}
\caption{\label{Fig14} Left: Schematic representation of an electrostatically tunable double-membrane PhC cavity showing the field distributions of symmetric (s) and antisymmetric (as) mode. Right: PL of the antisymmetric L3 mode of a double-membrane PhC cavity as function of the DC bias. The L3 mode gradually shifts in resonance with the QD lines (by sweeping the bias from 0 to -4.8 V. Reprinted with permission from \cite{Midolo2012}. \textcopyright 2012, AIP Publishing LLC.}
\end{figure}

Whereas a single exciton can easily be tuned into resonance with a cavity mode using the Stark effect, more elaborated integration schemes such as the interference of two indistinguishable single photons from two QDs in two separate photonic surroundings require the ability to control both the QD and cavity frequencies independent from each other so that different QD and cavity lines can be all brought to resonance \cite{Petruzzella2015}. In this regard, Stark tuning only affects the QD exciton energy and does not shift the cavity resonance. The separate tuning of PhC cavity resonances is a challenging task, especially at the low temperatures of interest. Several methods for cavity tuning have been proposed, including the use of photosensitive materials \cite{Faraon2008}, wet-chemical etching \cite{Hennessy2005}, nano-oxidation techniques \cite{Hennessy2006,Lee2009,Intonti2012,Piggott2014}, the insertion of gases \cite{Mosor2005} or liquids \cite{Intonti2009,Vignolini2010,Speijcken2012} as well as near-field probes \cite{Koenderink2005,Vignolini2008}. However, all these techniques are not very convenient in the prospect of realizing fully integrated quantum photonic circuits. None of these approaches allows the real-time tuning of each cavity in an array, as needed for QPICs. The latter functionality can be achieved using nano-electro-mechanical actuation, for example using a double-membrane cavity \cite{Notomi2006} (see Figure \ref{Fig14}). This was recently demonstrated in \cite{Midolo2011}. In this structure, the PhC mode extends over two closely-spaced membranes, and an electromechanical control of their distance results in a change of the mode wavelength, potentially over several tens of nm. With this structure, electrical tuning of the PhC resonance over $>$10\,nm at 10\,K was obtained \cite{Midolo2012} (see Figure \ref{Fig14}), without affecting the QD emission lines. Combining this cavity tuning with the Stark tuning of the QD excitons results in widely tunable single-photon sources to be used as key building blocks for scalable integrated photonic quantum processing \cite{Petruzzella2015}. Note that those electromechanical PhC cavities can also be used as tunable filters - a much needed component in quantum photonic integrated circuits with detectors.

\begin{figure}
\center
\includegraphics[width=0.8\linewidth]{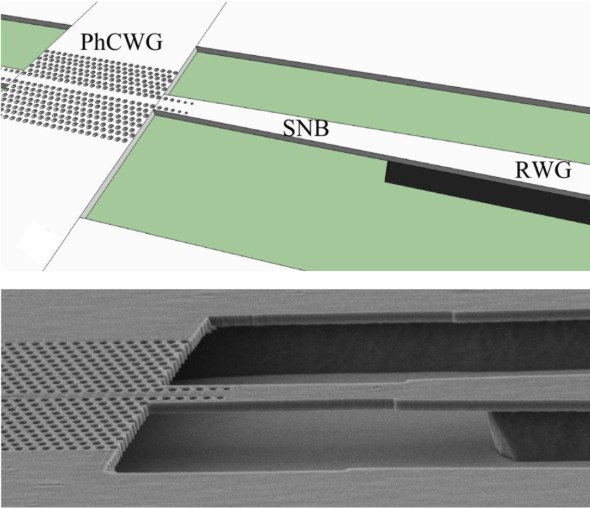}
\caption{\label{Fig15} Top: sketch of the interface between PhC waveguide, suspended nanobeam and ridge waveguide. Bottom: scanning electron microscopic image of the fabricated waveguide devices at the interface of suspended and ridge waveguide. Reprinted with permission from \cite{Fattahpoor2013}. \textcopyright 2013, AIP Publishing LLC.}
\end{figure}

\section{Photon Routing and Manipulation}
\subsection{Photon transport in ridge waveguides}
While PhC WGs are attractive for the modification of the LDOS and the optimization of the efficiency, a photonic circuit fully based on the photonic crystal geometry typically suffers from high optical losses (although losses down to 8\,dB/cm have been demonstrated on GaAs \cite{Sugimoto2004}) and substantial structural instability. A more efficient way of transporting single photons through photonic circuits is the use of ridge waveguides (RWGs) as they exhibit good mode confinement and low losses ($< 1$\,dB/cm \cite{Sugimoto2004}). This requires the efficient coupling of single photons from PhC waveguides to low-loss, supported ridge waveguides. Coupling from the PhC waveguide into the RWG using a simple, single-step lithographic process was recently demonstrated showing coupling efficiencies of up to 70\% \cite{Fattahpoor2013}. It is based on tapering the mode in both the lateral and the vertical direction by gradually changing the width of the waveguide (see Figure \ref{Fig15}). By using this structure, single photons were coupled to a ridge waveguide at a rate as high as 3.5\,MHz, which exceeds the typical rates obtained by coupling single-photon emission from QDs into fibers, showing the huge advantage of integration.

\subsection{Optical phase shifters}
Phase tuning and/or photon switching is needed in many implementations of photonic QIP, such as boson sampling \cite{Spring2013}. Additionally, fine tuning of the splitting ratio of integrated beamsplitters is often required to correct fabrication imperfections. These circumstances require a robust way to manipulate the phase of photons in a given waveguide of an individual input arm. Typically, this is achieved by using electro- or acousto-optical modulators which are based on external modulations of the refractive index of the medium in which the single photons propagate.

Nowadays, electro-optic phase modulators (EOPMs) are routinely fabricated and operated by applying a voltage across a non-centrosymmetric material (e.g. GaAs) vertical to the propagation direction through metallic electrodes. The voltage causes a change of the refractive index as function of the applied electric field. In general, the quality of an EOPM is related to $V_\pi L$, the voltage length product required for a phase shift of $\pi$, and should ideally be $\ll$1\,Vcm to enable dense packaging of photonic components. GaAs presents an enhancement of the electro-optic effect at energies very close to the bandgap \cite{Stievater2009}. In this wavelength region it has a very large electro-optic coefficient ($\gamma = 2.4\times10^{-11}$\,m/V \cite{Tatebayashi2007}) and relatively low values of $V_{\pi}L$ = 0.21\,Vcm \cite{Shin2008}. 

Suspended GaAs waveguides can confine the optical mode very tightly due to the high index contrast between GaAs and air. By interconnecting the waveguide with bridges for electrical connection, only very small voltages are necessary to achieve large electric field strengths as result of the short inter-electrode distance. An effective doping scheme with gradually doped $p$- and $n$-layers can further reduce $V_\pi$ \cite{Shin2008}.

A different way of changing the optical phase of a photon wave is the structural modification of the waveguide. This can be achieved for example through the application of a surface acoustic wave via a piezoelectric transducer perpendicular to the propagation direction of photons. The acoustic wave generates regions of compression and extraction. This changes the refractive index of the waveguide material via the elasto- and electro-optic effect due to the applied strain and the inherent piezoelectricity of GaAs, respectively. These acousto-optic modulations can operate up to GHz frequencies \cite{deLima2006}. However, those transducers are hard to implement, require substantial power and cannot be used to fix the phase over long time scales.

A very promising alternative for the phase tuning and reconfiguration of QPICs is given by the use of low-frequency micro- and nano-mechanical structures, where a physical displacement is produced via electrical actuation (e.g. through capacitive forces). Nanomechanical phase modulators have been demonstrated, for example in silicon \cite{Poot2014}. The double-membrane structure used for the tuning of PhC cavities\cite{Midolo2011,Midolo2012} lends itself to this type of phase modulators, potentially featuring $V_{\pi}L$ products in the order of 0.005 Vcm and maximum frequencies in the MHz range.

\begin{figure}
\center
\includegraphics[width=0.95\linewidth]{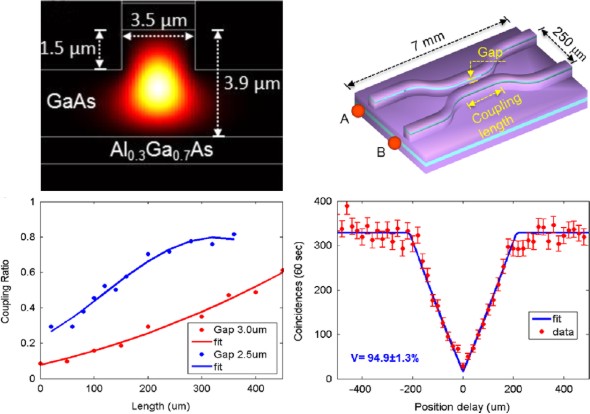}
\caption{\label{Fig16} Top left: Cross section of a GaAs/AlGaAs ridge waveguide and the simulated field distribution of the fundamental TE-mode. Top right: Schematic image of a GaAs directional coupler. Bottom left: Coupling ratio of the couplers for different gaps as function of coupling length. Bottom right: Two-photon quantum interference in the directional couplers with a coupling ratio $\varepsilon \approx 0.5$ showing high visibility of almost 95\%. Reprinted from \cite{Wang2014} , \textcopyright 2014, with permission from Elsevier.}
\end{figure}

\subsection{Beam splitters, directional couplers and interferometers}
Quantum integrated photonic circuits not only provide a platform to monolithically integrate single-photon sources, large-scale quantum circuits and detectors, but also offer high-visibility classical and quantum interference due to the inherent stability of the interferometers and perfect mode-overlap at the beam splitters. Using dual-rail (or path) encoding, an arbitrary single path-encoded qubit can be represented as a superposition of two states: $\alpha\left|10\right\rangle+\beta\left|01\right\rangle$ (in this notation the state $\left| nm\right\rangle$ indicates the presence of $n$ photons in one waveguide and $m$ in the other). An integrated optical beam splitter, usually implemented with a 50:50 directional coupler or a multimode interference (MMI) coupler, can easily perform a Hadamard operation and produce a superposition state such as ($\left|10\right\rangle+\left|01\right\rangle)/\sqrt{2}$ from a single photon propagating in one waveguide. Moreover, when two indistinguishable photons meet at the Hadamard gate, quantum interference produces photon bunching and a maximally entangled state (both photons are either reflected or transmitted). The initial input state $\left|11\right\rangle$ is then transformed into the entangled state ($\left|20\right\rangle+\left|02\right\rangle)/\sqrt{2}$ after the Hadamard operation. The same quantum interference, in combination with detection, is also at the origin of the photon non-linearity exploited in linear optics quantum computing. 

\begin{figure*}
\center
\includegraphics[width=0.66\linewidth]{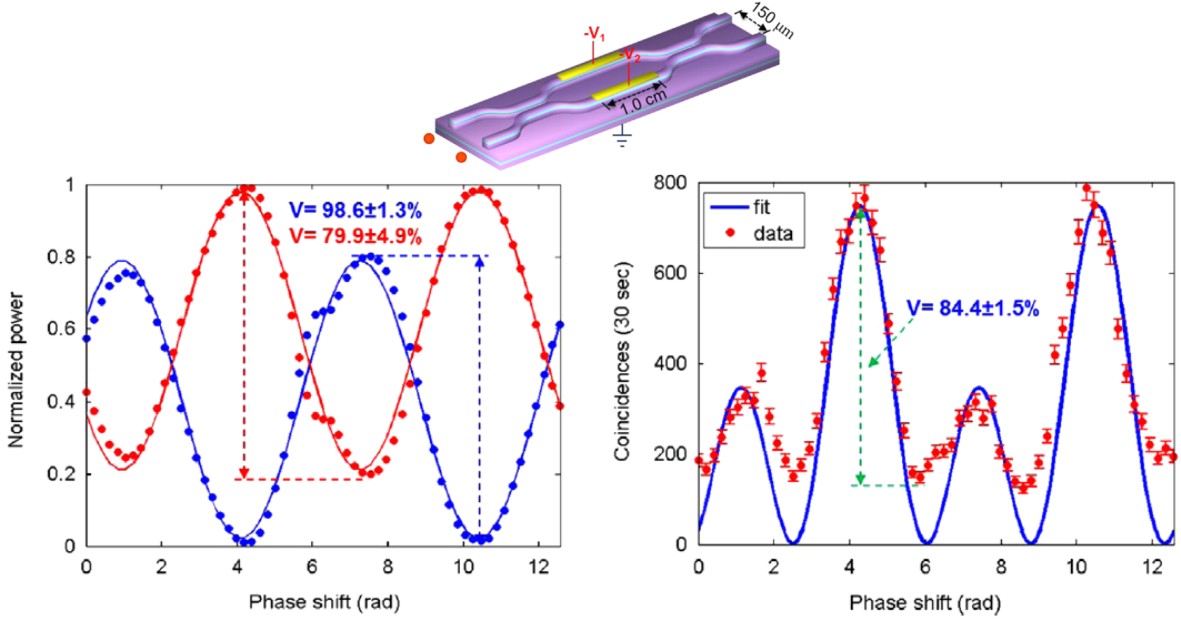}
\caption{\label{Fig17} Top left: Schematic image of the Mach-Zehnder interferometer (MZI) with two directional couplers and two electro-optical phase shifters. Top right: Optical image of the fabricated MZIs. Bottom left: Normalized intensities of the two outputs as function of relative phase shift for coherent bright-light input (for a coupling ratio of $\varepsilon = 0.3$). Bottom right: Quantum interference fringes showing manipulation of the two-photon state. Reprinted from \cite{Wang2014}, \textcopyright 2014, with permission from Elsevier.}
\end{figure*}

This key effect was already demonstrated in GaAs waveguides \cite{Wang2014,Joens2015} where directional couplers were designed and fabricated using GaAs/AlGaAs ridge waveguides with widths and heights of 2-4\,$\mu$m. Figure \ref{Fig16} shows the calculated field distribution for the quasi-TE fundamental mode and schematic image of the fabricated GaAs/AlGaAs directional couplers. A propagation loss of 2.5-3.5\,dB/cm and a coupling loss of 2-2.5\,dB/facet were measured using the Fabry-P\'{e}rot method. The coupling ratio $\varepsilon$ of the directional coupler is defined as the ratio of reflectivity and transmission. By varying the length of the coupler, an arbitrary value for $\varepsilon$ can be set. As can be seen in Figure \ref{Fig16}, a near 50:50 directional coupler with two bent input/output waveguides can be realized when the coupling length is about 140\,$\mu$m for a 2.5\,$\mu$m gap. Single photons pairs were launched into the 50:50 directional coupler and the coincidence counts after the coupler were measured using a counting module and two single-photon detectors. By changing the time arrival of the two photons, a Hong-Ou-Mandel (HOM) dip with a high visibility of almost 95\% was observed as shown in Figure \ref{Fig16}. On-chip beamsplitters were recently integrated with single-photon sources based on QDs \cite{Prtljaga2014,Joens2015,Rengstl2015}.

Besides the creation of entangled states by using directional couplers, also arbitrary unitary operations of quantum states are required to implement quantum communication and universal quantum computing. In order to prepare and measure arbitrary path-encoded qubits, two Hadamard gates and three phase shifters are required (representing a Mach-Zehnder interferometer), one of which induces a relative phase and amplitude between the two waveguides (see Figure \ref{Fig17}). The Mach-Zehnder interferometer then transforms a single-photon quantum state $\left|10\right\rangle$ into $\left[(1-2\varepsilon)\cos(\theta/2)+\imath\sin(\theta/2)\right]\left|10\right\rangle + 2\imath\sqrt{\varepsilon(1-\varepsilon)}\cos(\theta/2)\left|01\right\rangle$ where $\theta$ is the relative phase between the two arms \cite{Wang2014}. Thus, a simple integrated Mach-Zehnder interferometer (MZI) with two 50:50 directional couplers and three phase shifters can prepare and manipulate an arbitrary quantum state. 

GaAs/AlGaAs MZIs consisting of two identical 50:50 directional couplers and two electro-optical phase shifters on the top of two arms were demonstrated in \cite{Wang2014}, a schematic representation and an optical microscopic image can be seen in Figure \ref{Fig17}. The working principle was demonstrated by using both a classical light source and a single photon source. Classical light showed sinusoidal oscillations at both outputs. Their obvious unbalance was ascribed to the non-perfect coupling ratio of $\varepsilon$ = 0.3. When two single photons are separately launched into the two input ports of the Mach-Zehnder interferometer, the first coupler creates the two-photon state $\sqrt{2\varepsilon(1-\varepsilon)}\imath(\left|20\right\rangle+\left|02\right\rangle)$ + $(1-2\varepsilon)\left|11\right\rangle$. For $\varepsilon = 0.5$, the two photons are maximally path-entangled. The phase shifters then perform a rotation and the second coupler transforms the state to
\begin{eqnarray*}
\imath\sqrt{2\varepsilon(1-\varepsilon)}\left[-\varepsilon e^{-2\imath\theta}+(1-2\varepsilon) e^{-\imath\theta}+1-\varepsilon\right]\left|20\right\rangle\\
+\imath\sqrt{2\varepsilon(1-\varepsilon)}\left[(1-\varepsilon) e^{-2\imath\theta}+(1-2\varepsilon) e^{-\imath\theta}-\varepsilon\right]\left|02\right\rangle\\
+\left[-2\varepsilon(1-\varepsilon) e^{-2\imath\theta}+(1-2\varepsilon)^2 e^{-\imath\theta}-2\varepsilon(1-\varepsilon)\right]\left|11\right\rangle
\end{eqnarray*}
The manipulation of the two-photon entanglement state can be seen in Figure \ref{Fig17} showing two-photon quantum interference with a doubled frequency compared to the classical light experiment.

\subsection{Switching and routing}
There are several ways of controlling the propagation of single photons by using the inherent optical properties of quantum dots. The two most prominent examples herein are single-photon switching and spin-photon routing.

Single-photon switches are systems based on nonlinear optical interactions \cite{Kasprzak2010,Loo2012} in which the injection of control photons strongly affects the propagating of signal photons. This effect was already observed in strongly coupled QD-cavity systems (where strong coupling sets in when the coherent coupling constant of the QD-cavity systems becomes larger than the photon decay rate)\cite{Volz2012,Englund2012,Bose2012}. In this case, the energy structure, described by the Jaynes-Cummings ladder, is anharmonic, which means that, by exciting the system at the frequency of the transition to the upper polariton state in the first manifold with a control photon, the scattering rate of photons in the second manifold is changed drastically \cite{Volz2012}. Typical switching times lie in the ps-range. Single-photon switches thus may represent a key element in future high-bandwidth QPICs.

Another way of externally manipulating the propagation of single photons is given by the exploitation of the QD spin degree of freedom. Spin states such as those in a semiconductor QD are considered to be very good qubits \cite{Warburton2013} as they possess coherence times in the tens of nanoseconds range, which can be extended to the microseconds range by spin-echo techniques \cite{Petta2005}. The spin of resident electrons in QDs can be mapped onto the circular polarization state of photons emitted by the related charged exciton (trion) \cite{Berezovsky2008}, when an additional exciton is pumped in the QD \cite{Atature2007}. However, this in-plane circular polarization state is difficult to map into polarization- or path-encoded photons propagating along the layers. Additionally, the strong birefringence of nanophotonic waveguides makes the transmission of polarization-encoded qubits difficult.

\begin{figure}
\center
\includegraphics[width=0.95\linewidth]{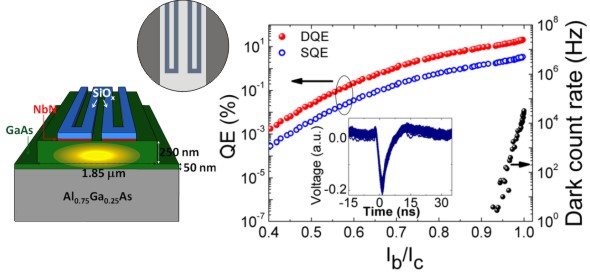}
\caption{\label{Fig18}Left: SEM image of perpendicular waveguides with grating outcouplers at each end. The QD is located at an off-center position (90 nm away from the center) as indicated in the inset. Right: PL spectra from the off-center QD excited with circularly polarised light and recorded at opposite grating outcouplers for magnetic fields between $\pm 4$\,T. Reprinted with permission from \cite{Luxmoore2013b}. \textcopyright 2011, AIP Publishing LLC.}
\end{figure}

This difficulty can be circumvented by a proper design of the electromagnetic environment around the QD. For example, by placing a QD at the crossing point of two orthogonally arranged photonic waveguides (Figure \ref{Fig18}, left) and detecting the output at the waveguide ends, the state of a spin-polarized exciton can be mapped onto the propagation direction of the generated photons \cite{Luxmoore2013a}. Although this experiment is hardly compatible with the KLM scheme in terms of scalability, its outcome has triggered the idea of using QD spin states to transfer vertically coupled photon polarization degrees into in-plane path qubits. It was found that, by placing the QD off-center (see Figure \ref{Fig18}), the individual spin polarization can deterministically be addressed (and directed within the photonic circuit) by using circularly polarized light \cite{Luxmoore2013b}. In the same way, when placing a QD into the so-called $C$-point of a W1 PhC waveguide, in principle, unidirectional emission and even an entangled photon source can be achieved \cite{Young2015}. Recently, this scheme was experimentally demonstrated in a glide-plane waveguide whose mirror-symmetry is broken by gradually shifting the inner hole row away from the W1 defect (towards one end of the waveguide) leading to a preferential emission of photons from a circularly polarized dipole into one of the two directions (depending on the helicity) \cite{Soellner2015}.

\section{Single-photon detectors}
\subsection{Superconducting nanowire detectors}
\begin{figure}
\center
\includegraphics[width=0.99\linewidth]{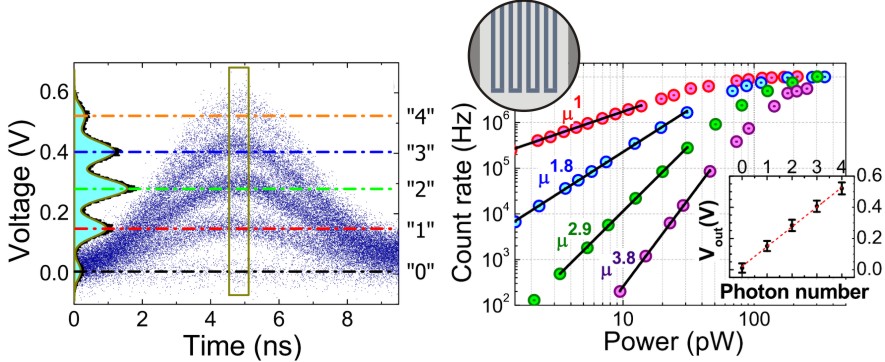}
\caption{\label{Fig19} Left: Sketch of a waveguide superconducting single-photon detector. Right: Device quantum efficiency (red dots, left axis) and dark count rate (blue dots, right axis) as a function of the normalized bias current. Inset: Output electrical pulse, showing a 1/$e$ time constant of 3.6\,ns. Reprinted with permission from \cite{Sprengers2011}. \textcopyright 2011, AIP Publishing LLC.}
\end{figure}

The integration of single-photon detectors (SPDs) with quantum photonic integrated circuits is particularly challenging, as the complex device structures associated with conventional single-photon detectors, such as avalanche photodiodes, are not easily compatible with the integration with low-loss waveguides and even less with sources. Transition-edge sensors (TES) may be suited for integration \cite{Gerrits2011}, but they are plagued by very slow response times (typically leading to maximum counting rates in the tens of kHz range) and require cooling down to $<100$\,mK temperatures. Instead, the "hot-spot" detection mechanism in superconducting nanowires \cite{Gol'tsman2001,Hadfield2009,Natarajan2012} is much more promising for fast single-photon detection. This detection mode can be implemented in a GaAs photonic circuit by sputtering an ultrathin (4-5\,nm) NbN film on top of GaAs/AlGaAs waveguide heterostructures and by its subsequent patterning into narrow wires\cite{Gaggero2010,Sprengers2011}. The evanescent field of the guided mode interacts with the wires (see Figure \ref{Fig19}a), resulting in a modal absorption coefficient of several hundreds of cm$^{-1}$, ensuring nearly 100\% absorptance in a propagation distance of a few tens of $\mu$m. 

The devices need to be cooled down to cryogenic temperatures and biased with a current $I_\text{b}$ close to the critical current $I_\text{c}$ of the superconducting wire. When the photon is absorbed, a region with a high concentration of quasi-particles is formed ("hot-spot"), leading to a local disruption of the superconductivity and resulting in a resistance appearing in the wire through a process involving the creation and crossing of vortices \cite{Renema2014,Renema2015}. The bias current is correspondingly expelled from the wire to the load formed by the input resistance of an amplifier producing a voltage pulse. The typical shape of such an output pulse is shown in Figure \ref{Fig19}, revealing a time constant being related to the recovery of the bias current in the wire after superconductivity has been re-established. This time constant (typically in the few ns range) is limited by the wire's kinetic inductance \cite{Kerman2006}. Maximum counting rates in the 100\,MHz range can be obtained, orders of magnitude higher than with TES detectors. The jitter in the output pulse, measured to be $~$ 60\,ps in waveguide SPDs on GaAs \cite{Sprengers2011} and even shorter in SPDs on Si \cite{Pernice2012}, is also much better than in TES and avalanche photodiodes, allowing much higher temporal resolution in single-photon measurements.

The quantum efficiency (QE) of superconducting SPDs is observed to strongly depend on the bias current, with a maximum close to the critical current, since the probability of switching the wire from the superconducting to the resistive state is maximum there. Besides that, also the nanowire geometry \cite{Marsili2011} as well as fabrication imperfections such as nanowire width and film thickness fluctuations \cite{Gaudio2014} strongly affect the quantum efficiency. The QE is further a function of the position across the nanowire width at which the incident photon is absorped \cite{Renema2015}. In GaAs-based waveguide SPDs, maximum device quantum efficiencies of about 20\% (defined as the photocounts normalized to the number of photons coupled into the WG for the transverse-electric (TE) polarization) were measured \cite{Sprengers2011}, which are promising for applications in quantum photonic integrated circuits. Note that even higher efficiencies were obtained on Si/SiO$_2$ \cite{Pernice2012,Calkins2013,Ferrari2015} or SiN$_x$ \cite{Kahl2015,Najafi2015} due to the possibility to deposit NbN at higher temperatures. Improvement of the quality of NbN films on GaAs, the use of lower-gap superconductors such as WSi \cite{Baek2011,Marsili2013} and the use of suspended or PhC GaAs waveguides \cite{Sahin2015} should result in QEs approaching 100\%, as needed in large-scale quantum photonic integrated circuits. 

The yield of SPDs on a chip is also a key issue for the scalability of integrated circuits. NbN-based SPDs are typically plagued by the low yield of efficient devices, which has been attributed to the presence of localized defects (constrictions) \cite{Kerman2007} and continuously distributed inhomogeneities \cite{Gaudio2014}. These regions with lower local critical current prevent a uniform biasing of the detector and thereby limit the efficiency. Their effect is expected to be less critical in WSi films due to the larger hot-spot diameter \cite{Baek2011} and the correspondingly lower dependence on the bias current. 

\subsection{Advanced functionalities with integrated detectors}
The waveguide configuration provides an opportunity to integrate additional functionality with the detector. For example, in \cite{Pernice2012}, an optical delay line constituted by a microring resonator was integrated with a nanowire detector on a Si waveguide. Nanophotonic cavities were integrated with superconducting nanowires, enabling spectral selectivity and near-unity efficiency \cite{Akhlaghi2015}. Superconducting nanowire detectors were also integrated with Si-based small-scale quantum photonic circuits \cite{Najafi2015}. In the GaAs platform, multimode interference (MMI) couplers, Mach-Zehnder interferometers and modulators can easily be integrated with detectors.

The detector structure itself can also be tailored to perform advanced measurements such as measuring the second-order auto-correlation function by integrating several wires on a waveguide \cite{Sahin2013a}. In this case, two NbN nanowires sense the electric field of the same guided mode. The nanowires are separately connected to two different bias and amplification circuits, providing two distinct read-outs of the photo-response signals, similar to the free-space configuration reported in \cite{Dauler2009}. By combining the detector outputs in a correlation card, the second-order correlation function of light propagating in the waveguide can be measured directly \cite{Sahin2013a}. This integrated auto-correlator provides the advantage of a very small footprint and a reduction in the total length of NbN wires needed to achieve a given absorption probability. Despite the very close physical proximity (wire-to-wire distance of 150\,nm), no crosstalk (spurious switching of one wire after detection of a photon in the other wire) was observed, within the experimental accuracy \cite{Sahin2013a}.

The on-chip detection of QD emission embedded in a ridge GaAs waveguide was recently demonstrated \cite{Reithmaier2013a,Reithmaier2015}. A major challenge in this regard is the spectral filtering of single QD emission lines and the discrimination of scattered laser light which can be achieved by pumping from the vertical direction and temporally filtering the QD emission from the short pump pulse \cite{Reithmaier2015}. 

\subsection{Photon-number resolving detectors}
\begin{figure*}
\center
\includegraphics[width=0.66\linewidth]{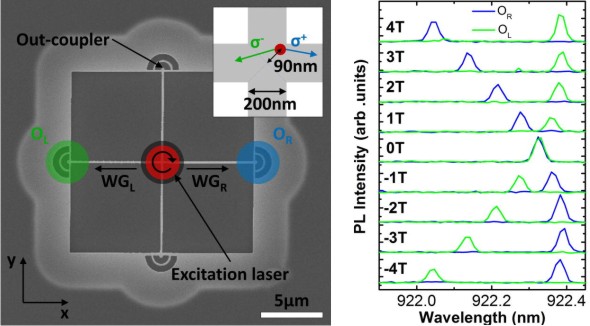}
\caption{\label{Fig20} Left: Oscilloscope persistence map of a photon flux detected with a waveguide photon-number resolving detector (consisting of four superconducting nanowires) showing 0-4 photon events. The left axis shows a corresponding histogram collected in a time frame indicated by the rectangle. Right: Count rates of the waveguide photon-number resolving detector corresponding to different photon counting levels: 1-photon (red), 2-photon (blue), 3-photon (green), and 4-photon (purple). Inset: the signal amplitude as a function of the detected photon number. Reprinted with permission from \cite{Sahin2013b}. \textcopyright 2011, AIP Publishing LLC.}
\end{figure*}

Apart from their use as autocorrelators, multi-wire structures can also be configured to work as photon-number-resolving (PNR) detectors\cite{Divochiy2008}. In this case, instead of reading out each wire separately, the wires are connected in parallel \cite{Divochiy2008} or in series \cite{Jahanmirinejad2012a,Jahanmirinejad2012b} in order to provide a single voltage output proportional to the number of switching wires. In particular, the series configuration has been shown to be attractive in view of the potential for high efficiency and scalability to large photon numbers\cite{Jahanmirinejad2012a,Zhou2014}. For example, as shown in Figure \ref{Fig20}, four wires can be patterned on top of the same waveguide, and connected in series, with an integrated resistor in parallel to each wire \cite{Sahin2013b}. When a wire switches to the resistive state after absorption of a guided photon, the bias current flowing through it is diverted to the parallel resistor, producing a voltage pulse. When two or more wires switch, the sum of the corresponding voltages is read out in the external circuit \cite{Sahin2013b}. Such waveguide PNR detectors will perform a fundamental function in future fully-integrated quantum photonic circuits, enabling photon number measurements with high efficiency in an extremely compact footprint.

\section{Integration perspectives and outlook}
\begin{figure}
\center
\includegraphics[width=0.99\linewidth]{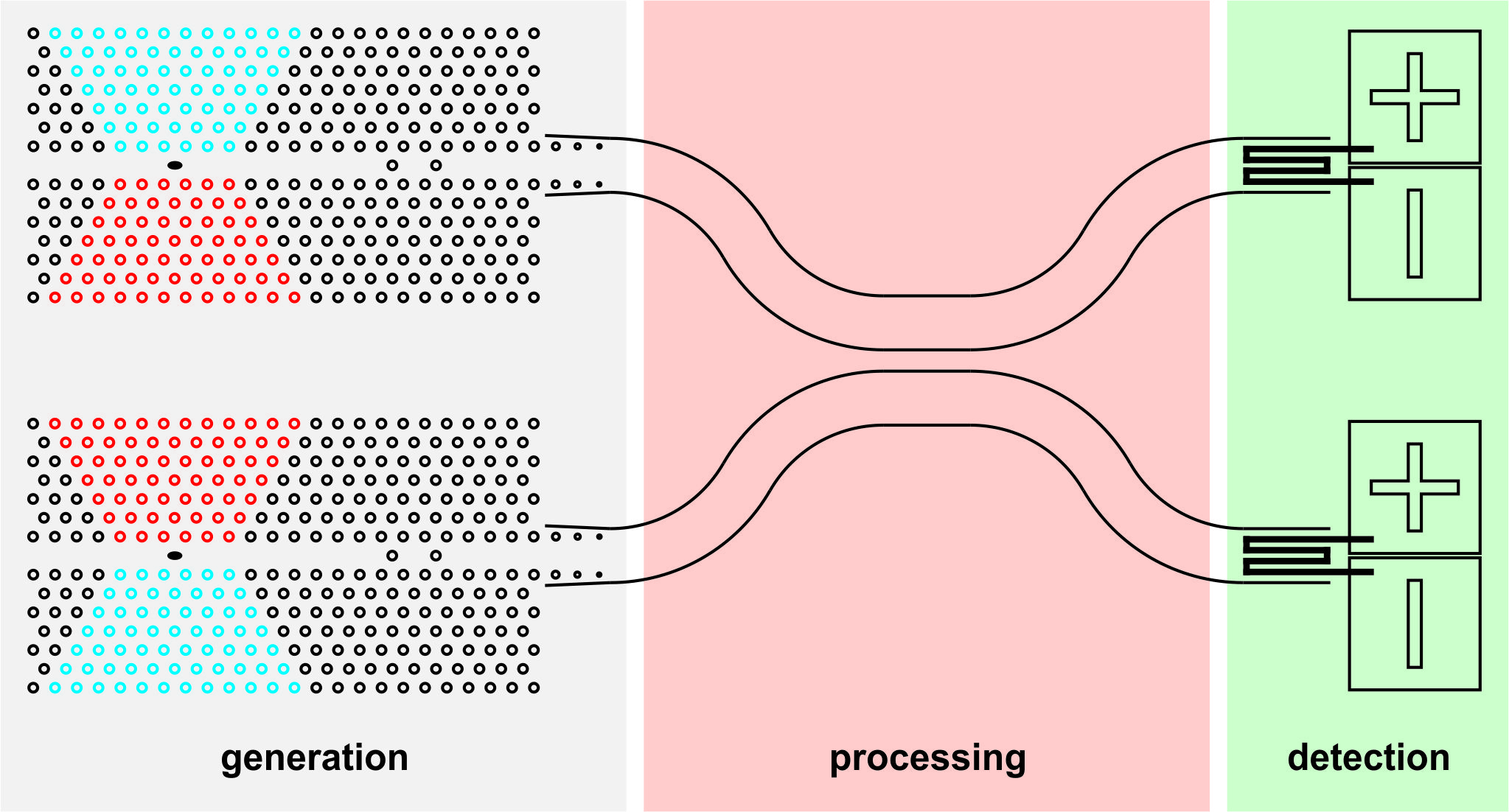}
\caption{\label{Fig21} Schematic illustration of a simple quantum photonic integrated circuit able to perform a second-order auto-correlation measurement and to demonstrate two-photon interference of single photons that are generated (by Purcell-enhanced emission from a QD in a PhC waveguide), filtered (by the PhC cavity), processed (by the two directional couplers) and detected (by the two superconducting nanowire detectors) on the same chip.}
\end{figure}

Figure \ref{Fig21} shows an illustration of a simple, possible quantum photonic integrated circuit, which enables to perform a two-photon interference measurement. In this circuit, single photons are generated by the radiative recombination of QD excitons placed in a PhC waveguide. The QD emission is tunable via the Stark effect created by the introduction of $p$- and $n$-doped regions into the PhC. The PhC waveguide transports the single photons towards ridge waveguides after being spectrally filtered by a PhC cavity in order to isolate photons from individual QD transition lines. The spectral filter itself can be tuned via a second underlying membrane (not shown). The single photons are then transferred into the beam splitter (directional couplers) where they exit in bunched pairs as result of their indistinguishability. The superconducting nanowire single photon detectors subsequently measure the second-order correlation function of the beam splitter output to prove the interference of single photons from two remote sources. This scheme can easily be extended to perform measurements of higher complexity by simply increasing the amount of individual units (sources, splitters and detectors) if a high efficiency is obtained in all parts of the circuit. In this way, multi-photon experiments such as boson sampling or advanced quantum photonic gates (e.g. reconfigurable chips\cite{Shadbolt2012}) can be integrated and scaled to large photon numbers.

All the building blocks for a multi-functional QPIC (such as illustrated in Fig.\ref{Fig21}), including single-photon sources, detectors and passive circuits, have been demonstrated on the GaAs platform. The corresponding efficiencies, while still below the values ($>$99\%) needed for scaling to many qubits, are already attractive as compared to the combination of coupling loss and efficiency typically affecting free-space or fiber implementations. Taking the case of a QD-based single-photon source at telecom wavelength as an example, the overall source-detector efficiency is in the $5\times10^{-4}$ range when using conventional micro-photoluminescence systems and commercial fiber-coupled SSPDs (resulting from a $~5\times10^{-3}$ probability per pulse of generating a single photon into a single-mode fiber and $10^{-1}$ detector quantum efficiency) \cite{Zinoni2007}, although higher efficiencies can be achieved for both the fiber-coupled source (up to $6\times10^{-2}$ \cite{Takemoto2010}) and the detector (up to 0.93 \cite{Marsili2013}). In the very first demonstrations of ridge-waveguide coupled devices, a photon generation probability of $6\times 10^{-2}$ per pulse \cite{Fattahpoor2013} and a detection quantum efficiency of 0.2 \cite{Sprengers2011} were reported, corresponding to a potential overall efficiency in the $10^{-2}$ range if the sources and detectors can be coupled efficiently on the same chip. 

We note that nearly the same waveguide epitaxial design was used for both the source and the detector, making their efficient coupling possible. Significant improvements in both integrated sources and detectors are possible, for example using nearly or strictly resonant excitation of the QD \cite{Takemoto2010,He2013b}, optimized PhC cavity design and processing, and improved superconducting nanowire technology, potentially bringing the overall efficiency in the tens of percent range. Additionally, integrating a large number of such devices on a single chip is in principle straightforward. Passive waveguide circuits are readily implemented with GaAs ridge waveguides. While the quantum interference experiments reported in Section 5.2 were carried out using waveguides with larger cores to optimize coupling to fibers, similar structures can easily be designed and realized based on the tightly confined ridge waveguide design used for sources and detectors. Waveguide losses were estimated to be in the few cm$^{-1}$ range for the relatively unoptimized technology used in \cite{Fattahpoor2013} and could certainly be improved. Due to the compact size of the needed components, waveguide loss is not expected to represent a major limitation for experiments involving a few photons.

A few challenges still need to be addressed before GaAs QPICs can be used to perform multi-photon experiments beyond the state-of-the-art: 

(i) \textit{Integrated filtering and buffering}: Tunable filters are always part of table-top quantum photonic experiments based on QDs, as they are needed to discriminate the emission from single excitonic lines, particularly in the case of non-resonant or near-resonant excitation. Filters can be realised using PhC cavities, and can be integrated with either the source or the detector. They can be made tunable by using the same electromechanical actuation needed for tuning the source \cite{Midolo2012}. Their integration in QPICs is therefore in principle straightforward. Another much needed component for quantum photonic integrated circuits is a single-photon buffer that is able to store single photons without destroying their non-classical nature and entanglement. Single-photon buffers can be realized by utilizing the control of the group index in PhC waveguides \cite{Takesue2013}, although it is challenging to combine a buffering time larger than the photon temporal length and low loss.

(ii) \textit{Improving the yield of sources and detectors and reducing losses of photonic components}: Scaling QPICs for multi-photon experiments requires arrays of sources and detectors with high yield and photonic components with vanishing optical losses. Increasing the yield in detector fabrication relies on improvements in superconducting thin film technology, as discussed above, with good outlook for the future. Achieving a high yield in sources, besides a relatively straightforward integration of the tuning techniques discussed in Section 4.3, further requires the active control of the QD spatial position, in order to achieve efficient coupling with the optical mode of the PhC cavity. Promising results have been shown in the position control of both strained InAs/GaAs QDs \cite{Atkinson2008,Schneider2009,Schneider2012,Joens2013} and pyramidal GaAs/AlGaAs QDs \cite{Pelucchi2007} (as discussed in Section 2.2), making the perspective of their integration in tunable PhC structures realistic. In this context, photonic components (such as waveguides, couplers etc.) still suffer from high optical losses that currently make possible up-scaling schemes unfeasible. Advanced lithographic steps need to be implemented in order to reduce losses and be able to implement true functionalities.

(iii) \textit{Improving the coherence and photon indistinguishability}: A major limitation of solid-state single-photon sources is the dephasing introduced by the environment as described in Section 3.1 and can be classified into pure dephasing (phase fluctuations on timescales shorter than the exciton lifetime) and spectral wandering (variations of the QD energy levels induced by a fluctuating charge environment on timescales longer than the exciton lifetime but shorter than the experiment duration). Both processes affect the photon indistinguishability and therefore the interference visibility observed in photon bunching experiments \cite{Santori2002}. Recent results on near-unity indistinguishability of photons generated by a single QD using pulsed resonant fluorescence \cite{He2013b} and quantum interference of photons generated by two distinct QDs \cite{Flagg2009,Patel2010,Konthasinghe2012a} suggest that these limitations can be surmounted by resonant pumping techniques and a careful control of the QD environment. Another very interesting route to the suppression of spectral wandering is the real-time control of the QD energy through a feedback loop \cite{Prechtel2013}. In any case, coherence remains the most challenging issue for the application of QD sources to multi-photon experiments.

Further down the line, high-density integration on GaAs will require increasing the refractive index contrast to achieve smaller radii of curvature in waveguide bends. This can be done by replacing the conventional GaAs/AlGaAs waveguide structure by a GaAs/SiO$_2$ structure, presenting a similar index contrast as Si/SiO$_2$. The bonding of GaAs membranes on SiO$_2$-coated Si substrates can be used for that purpose and has been demonstrated \cite{Alexe2000}. In this perspective of large-scales QPICs, residual waveguide loss will represent an important parameter and will have to be investigated carefully. We also note that other III-V active layers and membranes (and particularly those lattice-matched with InP) may be used instead of GaAs, leveraging on the vast developments in InP classical photonics. However, the progress on QD single-photon sources on InP has so far been limited \cite{Buckley2012}, and it is unclear what the limits will be in terms of photon indistinguishability.

A comparison between GaAs and other quantum photonic technologies is useful. The LiNbO$_3$, Si and SiN platforms can also in principle provide the key functionalities of photon production, passive control and detection, while the integration of efficient sources on silica-based QPICs appears very challenging. In terms of integration density, the small index contrast in LiNbO$_3$ represents a major fundamental limitation, making its application to large-scale circuits involving tens of photons and thousands of components doubtful. Si and SiN therefore represent the strongest competitors to GaAs for integrated quantum photonics. These materials benefit from decades of investments and technology developments in the electronics industry, allow fabricating devices with high quality and reproducibility, and are in principle compatible with production in CMOS fabs. The index contrast in both systems is high, enabling high integration levels. Indeed, fast progress, particularly in quantum silicon photonics, has recently led to the demonstration of QPICs with tens to hundreds of components \cite{Silverstone2013,Harris2015}. However, fundamental challenges remain in the large scale integration of sources, detectors and low-power reconfigurable circuits on the Si and SiN platforms. On the one hand, single-photon sources in these materials can only be based on two-photon production via spontaneous four-wave-mixing from a pump laser and heralding of a non-vacuum state by detection of one of the photons. This approach is intrinsically limited in terms of efficiency (probability of generating a single photon) since the average photon number must be kept low to avoid multi-photon events. A potential solution is the multiplexing of several sources \cite{Midgall2002}, which implies a very significant increase in the number of components and puts stringent requirements on the loss of switches \cite{Bonneau2015,FrancisJones2015}. On the other hand, the integration of superconducting nanowire detectors implies additional technological challenges on the Si and SiN platform, related to the extreme filter performance required to suppress the pump used for photon production, the impossibility to use thermal phase tuning at low temperature, and the incompatibility of most superconducting materials with CMOS fabs. A solution to these problems might lie in the use of hybrid platforms such as III-V materials on silicon by direct heteroepitaxial growth on top of each other or by wafer bonding\cite{Wada1999}.

While it is unclear at this point which technology will be suited for large-scale QPICs, the physical properties of GaAs and other III-Vs, including the direct bandgap and the non-centrosymmetric crystal structure (enabling a linear electro-optic effect), provide them a very fundamental advantage which could become crucial in the long term.

\section{Conclusion}
We have provided an overview of the recent progress in quantum integrated photonic components and circuits based on the GaAs technology platform. All key functionalities, including single-photon sources and detectors, photon-number-resolving detectors, integrated autocorrelators and tunable Mach-Zehnder interferometers have been realized and tested, and on-chip photon-photon interference has been demonstrated. The remaining challenges to be addressed for scaling GaAs quantum photonic integrated circuits to the level of few tens of photons have also been discussed. These results lay the foundation of a fully-integrated quantum photonic technology, with potential applications in quantum simulation and computing.

\begin{acknowledgement}
We gratefully acknowledge many useful discussions with S. Fattah poor, T.B. Hoang, A. Gaggero, R. Leoni, F. Mattioli, L. Midolo, M. Petruzzella, D. Sahin, J.P. Sprengers, T. Xia and C. Schneider. We further acknowledge financial support from the European Commission within the FP7 project QUANTIP (Project No. 244026) and the Dutch Technology Foundation STW (Project No. 10380).
\end{acknowledgement}

\begin{biographies}
\authorbox{}{Christof P. Dietrich}{received his PhD degree in physics from the Graduate School BuildMona of University of Leipzig, Germany. After postdoc positions in The Netherlands (TU Eindhoven) and the United Kingdom (University of St Andrews), he is now group leader in the Technical Physics Department of the University of W\"{u}rzburg, Germany, where he conducts projects on integrated GaAs quantum photonic structures. His current research interests also include polariton physics and organic photonics.}
\authorbox{}{Andrea Fiore}{graduated in electronic engineering and physics from the University of Rome "La Sapienza" (Italy) and obtained the PhD degree in 1997 from Univ. of Paris XI (France). He has held appointments at the University of California at Santa Barbara (USA), the Institute of Photonics and Nanotechnology of the Italian National Research Council (Italy), the Ecole Polytechnique F\'{e}d\'{e}rale de Lausanne (Switzerland) and is now Professor of Nanophotonics at the Eindhoven University of Technology (The Netherlands). His research interests are in the field of quantum photonics and semiconductor nanophotonics.}
\authorbox{}{Mark Thompson}{is Professor of Quantum Photonics at the University of Bristol, UK. He received his M.Phys. degree in Physics from the University of Sheffield UK, and  Ph.D. degree in Electrical Engineering from the University of Cambridge UK, in 2000 and 2007 respectively. Previous appointments include Research Scientist with Bookham Technology Ltd. UK, a Research Fellow at the University of Cambridge and a Research Fellow at the Toshiba Corporate R\&D Centre, Japan. He joined the University of Bristol in 2008, and currently leads a team of researchers developing integrated quantum photonic technologies for applications in quantum communications and computation.}
\authorbox{}{Martin Kamp}{studied physics at the University of W\"{u}rzburg, Germany, and obtained a Master of Arts degree from Stony Brook University, USA. He received a PhD degree from the University of W\"{u}rzburg in 2003 for his research on laterally coupled distributed feedback lasers. Until 2014, he was interims chair for the Technische Physik for several years and head of the Gottfried-Landwehr Laboratory for Nanotechnology. Since 2014, he is head of the Wilhelm Conrad R\"{o}ntgen Center for Complex Material Systems. His research interests include electronic and photonic semiconductor structures, quantum information processing and interband cascade lasers.}
\authorbox{}{Sven H\"{o}fling}{obtained the diploma degree in Applied Physics from the University of Applied Science in Coburg and his PhD degree from the University of W\"{u}rzburg. He was with the Fraunhofer Institute of Applied Solid-State Physics (Freiburg, Germany). In 2013, he became Professor in Physics at the University of St Andrews (United Kingdom). Since 2015, he holds the Chair for Technical Physics at the University of W\"{u}rzburg and is head of the Gottfried-Landwehr Laboratory for Nanotechnology. His research interests include the design, fabrication, and characterization of low-dimensional electronic and photonic nanostructures, including quantum wells and quantum dots, organic semiconductors, high-quality factor microcavities, photonic crystal devices and semiconductor lasers.}
\end{biographies}


\begin{thebibliography}{00}
\footnotesize
\bibitem{OBrien2009} J.L. O'Brien, A Furusawa and J. Vu\v{c}kovi\'{c}, Nat. Photon. \textbf{3}, 687 (2009). 
\bibitem{Faraon2011} A. Faraon, A. Majumdar, D. Englund, E. Kim, M. Bajcsy, and J. Vu\v{c}kovi\'{c}, New J. Phys. \textbf{13}, 055025 (2011).
\bibitem{Lodahl2015} P. Lodahl, S. Mahmoodian, and S. Stobbe, Rev. Mod. Phys. \textbf{87}, 347 (2015).
\bibitem{Knill2001} E. Knill, R. Laflamme, and G.J. Milburn, Nature \textbf{409}, 46 (2001).
\bibitem{Raussendorf2003} R. Raussendorf, D.E. Browne, and H.J. Briegel, Phys. Rev. A \textbf{68}, 022312 (2003).
\bibitem{Aaronson2010} S. Aaronson and A. Arkhipov, arXiv:1011.3245 (2010).
\bibitem{OBrien2003} J.L. O'Brien, G.J. Pryde, A.G. White, T.C. Ralph, D. and Branning, Nature \textbf{426}, 264-267 (2003).
\bibitem{Lanyon2009} B.P. Lanyon, M. Barbieri, M.P. Almeida, T. Jennewein, T.C. Ralph, K.J. Resch, G.J. Pryde, J.L. O'Brien, A. Gilchrist, A.G. White, Nat. Phys. \textbf{5}, 134 (2009).
\bibitem{Lanyon2007} B.P. Lanyon, T.J. Weinhold, N.K. Langford, M. Barbieri, D.F.V. James, A. Gilchrist, and A.G. White, Phys. Rev. Lett. \textbf{99}, 250505 (2007).
\bibitem{Lu2007} C.-Y. Lu, D.E. Browne, T. Yang, and J.-W. Pan, Phys. Rev. Lett. \textbf{99}, 250504 (2007).
\bibitem{Tillmann2013} M. Tillmann, B. Daki\'{c}, R. Heilmann, S. Nolte, A. Szameit, and P. Walther, Nat. Photon. \textbf{7}, 540 (2013).
\bibitem{Crespi2013} A. Crespi, R. Osellame, R. Ramponi, V. Giovannetti, R. Fazio, L. Sansoni, F. De Nicola, F. Sciarrino, and P. Mataloni, Nat. Photon. \textbf{7}, 322 (2013).
\bibitem{Spring2013} J.B. Spring, B.J. Metcalf, P.C. Humphreys, W.S. Kolthammer, X.-M. Jin, M. Barbieri, A. Datta, N. Thomas-Peter, N.K. Langford, D. Kundys, J.C. Gates, B.J. Smith, P.G.R. Smith, I.A. Walmsley, Science \textbf{339}, 798 (2013).
\bibitem{Broome2013} M.A. Broome, A. Fedrizzi, S. Rahimi-Keshari, J. Dove, S. Aaronson, T.C. Ralph, and A.G. White, Science \textbf{339}, 794 (2013).
\bibitem{Politi2008} A. Politi, M.J. Cryan, J.G. Rarity, S. Yu, J.L. O'Brien, Science \textbf{320}, 646 (2008).
\bibitem{Carolan2015} J. Carolan, C. Harrold, C. Sparrow, E. Mart\'{i}n-L\'{o}pez, N.J. Russell, J.W. Silverstone, P.J. Shadbolt, N. Matsuda, M. Oguma, M. Itoh, G.D. Marshall, M.G. Thompson, J.C.F. Matthews, T. Hashimoto, J.L. O'Brien, and A. Laing, Science \textbf{349}, 711 (2015).
\bibitem{Marshall2009} G.D. Marshall, A. Politi, J.C.F. Matthews, P. Dekker, M. Ams, M.J. Withford, and J.L. O'Brien, Opt. Expr. \textbf{17}, 12546 (2009).
\bibitem{Xiong2011} C. Xiong, W. Pernice, K.K. Ryu, C. Schuck, K.Y. Fong, T. Palacios, and H.X. Tang, Opt. Expr. \textbf{19}, 10462 (2011).
\bibitem{Jin2014} H. Jin, F.M. Liu, P. Xu, J.L. Xia, M.L. Zhong, Y. Yuan, J.W. Zhou, Y.X. Gong, W. Wang, and S.N. Zhu, Phys. Rev. Lett. \textbf{113}, 103601 (2014).
\bibitem{Bonneau2015} D. Bonneau, G.J. Mendoza, J.L. O'Brien, and M.G. Thompson, New J. Phys. \textbf{17}, 043057 (2015).
\bibitem{Harris2015} N.C. Harris, G.R. Steinbrecher, J. Mower, Y. Lahini, M. Prabhu, T. Baehr-Jones, M. Hochberg, S. Lloyd, and D. Englund, arXiv:1507.03406 (2015).
\bibitem{Wang2014} J.W. Wang, A. Santamato, P. Jiang, D. Bonneau, E. Engin, J.W. Silverstone, M. Lermer, J. Beetz, M. Kamp, S. H\"{o}fling, M.G.Tanner, C.M. Natarajan, R.H. Hadfield, S.N. Dorenbos, V. Zwiller, J.L. O'Brien, M.G. Thompson, Opt. Commun. \textbf{327}, 49 (2014).
%\bibitem{Tanzilli2001} S. Tanzilli, H. De Riedmatten, W. Tittel, H. Zbinden, P. Baldi, M. De Micheli, D.B. Ostrowsky, and N. Gisin, Electr. Lett. \textbf{37}, 26 (2001).
%\bibitem{Takesue2008} H. Takesue, H. Fukuda, T. Tsuchizawa, T. Watanabe, K. Yamada, Y. Tokura, and S. Itabashi, Opt. Expr. \textbf{16}, 5721 (2008).
\bibitem{Englund2007b} D. Englund, A. Faraon, I. Fushman, N. Stoltz, P. Petroff, and J. Vu\v{c}kovi\'{c}, Nature \textbf{450}, 857 (2007).
\bibitem{Lund-Hansen2008} T. Lund-Hansen, S. Stobbe, B. Julsgaard, H. Thyrrestrup, T. S\"{u}nner, M. Kamp, A. Forchel, and P. Lodahl, Phys. Rev. Lett. \textbf{101}, 113903 (2008).
\bibitem{Schwagmann2011} A. Schwagmann, S. Kalliakos, I. Farrer, J.P. Griffiths, G.A.C. Jones, D.A. Ritchie, and A.J. Shields, Appl. Phys. Lett. \textbf{99}, 261108 (2011).
\bibitem{Laucht2012} A. Laucht, S. P\"{u}tz, T. G\"{u}nthner, N. Hauke, R. Saive, S. Frederick, M. Bichler, M.-C. Amann, A.W. Holleitner, M. Kaniber, and J.J. Finley, Phys. Rev. X \textbf{2}, 011014 (2012).
\bibitem{Hoang2012} T.B. Hoang, J. Beetz, L. Midolo, M. Skacel, M. Lermer, M. Kamp, S. H\"{o}fling, L. Balet, N. Chauvin, and A. Fiore, Appl. Phys. Lett. \textbf{100}, 061122 (2012).
\bibitem{Sprengers2011} J.P. Sprengers, A. Gaggero, D. Sahin, S. Jahanmirinejad, G. Frucci, F. Mattioli, R. Leoni, J. Beetz, M. Lermer, M. Kamp, S. H\"{o}fling, R. Sanjines, and A. Fiore, Appl. Phys. Lett. \textbf{99}, 181110 (2011).
\bibitem{Gerrits2011} T. Gerrits, N. Thomas-Peter, J.C. Gates, A.E. Lita, B.J. Metcalf, B. Calkins, N.A. Tomlin, A.E. Fox, A.L. Linares, J.B. Spring, N.K. Langford, R.P. Mirin, J.B. Smith, I.A. Walmsley, and S.W. Nam, Phys. Rev. A \textbf{84}, 060301 (2011).
\bibitem{Pernice2012} W.H.P. Pernice, C. Schuck, O. Minaeva, M. Li, G.N. Gol'tsman, A.V. Sergienko, and H.X. Tang, Nat. Comm. \textbf{3}, 1325 (2012).
\bibitem{Reithmaier2013b} G. Reithmaier, J. Senf, S. Lichtmannecker, T. Reichert, F. Flassig, A. Voss, R. Gross, and J.J. Finley, J. Appl. Phys. \textbf{113}, 143507 (2013).
\bibitem{Walker1989} R.G. Walker, Appl. Phys. Lett. \textbf{54}, 1613 (1989). 
\bibitem{Leo1999} G. Leo, V. Berger, C. Ow Yang, and J. Nagle, J. Opt. Soc. Am. B \textbf{16}, 1597 (1999).
\bibitem{Collins2013} M.J. Collins,	C. Xiong,	I.H. Rey,	T.D. Vo,	J. He,	S. Shahnia,	C. Reardon,	T.F. Krauss,	M.J. Steel,	A.S. Clark, and B.J. Eggleton, Nat. Commun. \textbf{4}, 2582 (2013).
\bibitem{Arcari2014} M. Arcari, I. S\"{o}llner, A. Javadi, S. Lindskov Hansen, S. Mahmoodian, J. Liu, H. Thyrrestrup, E.H. Lee, J.D. Song, S. Stobbe, and P. Lodahl, Phys. Rev. Lett. \textbf{113}, 093603 (2014).
\bibitem{Orieux2013} A. Orieux, A. Eckstein, A. Lemaître, P. Filloux, I. Favero, G. Leo, T. Coudreau, A. Keller, P. Milman, and S. Ducci, Phys. Rev. Lett. \textbf{110}, 160502 (2013). 
\bibitem{Boitier2014} F. Boitier, A. Orieux, C. Autebert, A. Lema\^{i}tre, E. Galopin, C. Manquest, C. Sirtori, I. Favero, G. Leo, and S. Ducci, Phys. Rev. Lett. \textbf{112}, 183901 (2014).
\bibitem{Autebert2015} C. Autebert, G. Boucher, F. Boitier, A. Eckstein, I. Favero, G. Leo, and S. Duccia, J. Mod. Opt. \textbf{62}, 1739 (2015).
\bibitem{Snyder1991} C. W. Snyder, B. G. Orr, D. Kessler, and L. M. Sander, Phys. Rev. Lett. \textbf{66}, 3032 (1991).
\bibitem{Leonard1993} D. Leonard, M. Krishnamurthy, C.M. Reaves, S.P. Denbaars, and P.M. Petroff, Appl. Phys. Lett. \textbf{63}, 3203 (1993).
\bibitem{Moison1994} J.M. Moison, F. Houzay, F. Barthe, L. Leprince, E. Andr\'{e}, and O. Vatel, Appl. Phys. Lett. \textbf{64}, 196 (1994).
\bibitem{Grundmann1995} M. Grundmann, N.N. Ledentsov, R. Heitz, L. Eckey, J. Christen, J. B\"{o}hrer, D. Bimberg, S.S. Ruvimov, P. Werner, U. Richter, J. Heydenreich, V.M. Ustinov, A.Yu. Egorov, A.E. Zhukov, P.S. Kopev, and Zh.I. Alferov, Phys. Stat. Sol. B \textbf{188}, 249 (1995).
\bibitem{Oshinowo1994} J. Oshinowo, M. Nishioka, S. Ishida, and Y. Arakawa, Appl. Phys. Lett. \textbf{65}, 1421 (1994).
\bibitem{Petroff1994} P.M. Petroff, and S.P. DenBaars, Superlattices Microstruct. \textbf{15}, 15 (1994).
\bibitem{Heinrichsdorff1996} F. Heinrichsdorff, A. Krost, M. Grundmann, D. Bimberg, A. Kosogov, and P. Werner, Appl. Phys. Lett. \textbf{68}, 3284 (1996).
\bibitem{Michler2000} P. Michler, A. Kiraz, C. Becher, W.V. Schoenfeld, P.M. Petroff, L. Zhang, E. Hu, and A. Imamoglu, Science \textbf{290}, 2282 (2000).
\bibitem{Santori2002} C. Santori, D. Fattal, J. Vu\v{c}kovi\'{c}, G.S. Solomon, and Y. Yamamoto, Nature \textbf{419}, 594 (2002).
\bibitem{Patel2010} R.B. Patel, A.J. Bennett, I. Farrer, C.A. Nicoll, D.A. Ritchie, and A.J. Shields, Nat. Photon. \textbf{4}, 632 (2010).
\bibitem{Akopian2006} N. Akopian, N.H. Lindner, E. Poem, Y. Berlatzky, J. Avron, D. Gershoni, B.D. Gerardot, and P.M. Petroff, Phys. Rev. Lett. \textbf{96}, 130501 (2006).
\bibitem{Reithmaier2004} J.P. Reithmaier, G. Sek, A. L\"{o}ffler, C. Hofmann, S. Kuhn, S. Reitzenstein, L.V. Keldysh, V.D. Kulakovskii, T. L. Reinecke, and A. Forchel, Nature \textbf{432}, 197 (2004).
\bibitem{Yoshie2004} T. Yoshie, A. Scherer, J. Hendrickson, G. Khitrova, H.M. Gibbs, G. Rupper, C. Ell, O.B. Shchekin, and D.G. Deppe, Nature \textbf{432}, 200 (2004).
\bibitem{Stranski1938} I.N. Stranski and L. Krastanov, Mathematische Naturwissenschaftliche Klasse 2B \textbf{146}, 79 (1938).
\bibitem{Eisele2008} H. Eisele, A. Lenz, R. Heitz, R. Timm, M. D\"{a}hne, Y. Temko, T. Suzuki, and K. Jacobi, J. Appl. Phys. \textbf{104}, 124301 (2008).
\bibitem{Loeffler2006} A. L\"{o}ffler, J.-P. Reithmaier, A. Forchel, A. Sauerwald, D. Peskes, T. K\"{u}mmell, G. Bacher, J. Cryst. Growth \textbf{286}, 6 (2006).
\bibitem{Wang2006} L. Wang, A. Rastelli, and O. G. Schmidt, J. Appl. Phys. \textbf{100}, 064313 (2006). 
\bibitem{Garcia1998} J.M. Garcia, T. Mankad, P.O. Holtz, P.J. Wellman, and P.M. Petroff, Appl. Phys. Lett. \textbf{72}, 3172 (1998).
\bibitem{Yang2007} T. Yang, J. Tatebayashi, K. Aoki, M. Nishioka, and Y. Arakawa, Appl. Phys. Lett. \textbf{90}, 111912 (2007).
\bibitem{Ellis2007} D.J.P. Ellis, A.J. Bennett, A.J. Shields, P. Atkinson, and D.A. Ritchie, Appl. Phys. Lett. \textbf{90}, 233514 (2007).
\bibitem{Nishi1999} K. Nishi, H. Saito, S. Sugou, and J.-S. Lee, Appl. Phys. Lett. \textbf{74}, 1111 (1999).
\bibitem{Yeh2000} N.-T. Yeh, T.-E. Nee, J.-I. Chyi, T.M. Hsu, and C.C. Huang, Appl. Phys. Lett. \textbf{76}, 1567 (2000).
\bibitem{Alloing2007} B. Alloing, C. Zinoni, L.H. Li, A. Fiore, and G. Patriarche, J. Appl. Phys. \textbf{101}, 024918 (2007).
\bibitem{Li2013} S.L. Li, Q.Q. Chen, S.C. Sun, Y.L. Li, Q.Z. Zhu, J.T. Li, X.H. Wang, J.B. Han, J.P. Zhang, C.Q. Chen, and Y.Y. Fang, Nanoscale Res. Lett. \textbf{8}, 367 (2013). 
\bibitem{Convertino2004} A. Convertino, L. Cerri, G. Leo, S. Viticoli, J. Cryst. Growth \textbf{261}, 458 (2004).
\bibitem{Huang2007} S. Huang, Z. Niu, H.Q. Ni, Y.H. Xiong, F. Zhan, Z. Fang, J.B. Xia, J. Cryst. Growth \textbf{301}, 751 (2007).
\bibitem{Koguchi1993} N. Koguchi and K. Ishige, Jpn. J. Appl. Phys. \textbf{32}, 2052 (1993).
\bibitem{Gammon1996} D. Gammon, E.S. Snow, B.V. Shanabrook, D.S. Katzer, and D. Park, Science \textbf{273}, 87 (1996).
\bibitem{Hours2005} J. Hours, P. Senellart, E. Peter, A. Cavanna, and J. Bloch, Phys. Rev. B \textbf{71}, 161306(R) (2005).
\bibitem{Mano2009} T. Mano, M. Abbarchi, T. Kuroda, C.A. Mastrandrea, A. Vinattieri, S. Sanguinetti, K. Sakoda, and M. Gurioli, Nanotechnology \textbf{20}, 395601 (2009).
\bibitem{Tighineanu2013} P. Tighineanu, R. Daveau, E.H. Lee, J.D. Song, S. Stobbe, and P. Lodahl, Phys. Rev. B \textbf{88}, 155320 (2013).
\bibitem{Peter2005} E. Peter, P. Senellart, D. Martrou, A. Lema\^{i}tre, J. Hours, J.M. G\'{e}rard, and J. Bloch, Phys. Rev. Lett. \textbf{95}, 067401 (2005).
\bibitem{Badolato2005} A. Badolato, K. Hennessy, M. Atat\"{u}re, J. Dreiser, E. Hu, P.M. Petroff, A. Imamoglu, Science \textbf{308}, 1158 (2005).
\bibitem{Hennessy2007} K. Hennessy, A. Badolato, M. Winger, D. Gerace, M. Atat\"{u}re, S. Gulde, S. F\"{a}lt, E.L. Hu, A. Imamoglu, Nature \textbf{445}, 896 (2007).
\bibitem{Suenner2008} T. S\"{u}nner, C. Schneider, M. Strau\ss, A. Huggenberger, D. Wiener, S. H\"{o}fling, M. Kamp, and A. Forchel, Opt. Lett. \textbf{33}, 1759 (2008).
\bibitem{Schmidt2007} O.G. Schmidt, \textit{Lateral Alignment of Epitaxial Quantum Dots} (Springer, Berlin, 2007).
\bibitem{Krenner2005} H.J. Krenner, M. Sabathil, E.C. Clark, A. Kress, D. Schuh, M. Bichler, G. Abstreiter, and J.J. Finley, Phys. Rev. Lett. \textbf{94}, 057402 (2005).
\bibitem{Heidemeyer2004} H. Heidemeyer, C. M\"{u}ller, O.G. Schmidt, J. Cryst. Growth \textbf{261}, 444 (2004).
\bibitem{Kiravittaya2006} S. Kiravittaya, M. Benyoucef, R. Zapf-Gottwick, A. Rastelli, and O. G. Schmidt, Appl. Phys. Lett. \textbf{89}, 233102 (2006).
\bibitem{Gallo2008} P. Gallo, M. Felici, B. Dwir, K. A. Atlasov, K. F. Karlsson, A. Rudra, A. Mohan, G. Biasiol, L. Sorba, and E. Kapon, Appl. Phys. Lett. \textbf{92}, 263101 (2008).
\bibitem{Mereni2009} L.O. Mereni, V. Dimastrodonato, R.J. Young, and E. Pelucchi, Appl. Phys. Lett. \textbf{94}, 223121 (2009).
\bibitem{Huggenberger2011a} A. Huggenberger, S. Heckelmann, C. Schneider, S. H\"{o}fling, S. Reitzenstein, L. Worschech, M. Kamp, and A. Forchel, Appl. Phys. Lett. \textbf{98}, 131104 (2011).
\bibitem{Helfrich2012} M. Helfrich, P. Schroth, D. Grigoriev, S. Lazarev, R. Felici, T. Slobodskyy, T. Baumbach, and D.M. Schaadt, Phys. Stat. Sol. A \textbf{209}, 2387 (2012).
\bibitem{Maier2014a} S. Maier, K. Berschneider, T. Steinl, A. Forchel, S. H\"{o}fling, C. Schneider, and M. Kamp, Semicond. Sci. Technol. \textbf{29}, 052001 (2014).
\bibitem{Unsleber2015b} S. Unsleber, S. Maier, D.P.S. McCutcheon, Y.-M. He, M. Dambach, M. Gschrey, N. Gregersen, J. M\o rk, S. Reitzenstein, S. H\"{o}fling, C. Schneider, and M. Kamp, Optica \textbf{2}, 1072 (2015).
\bibitem{MartinSanchez2009} J. Martin-Sanchez, G. Munoz-Matutano, J. Herranz, J. Canet-Ferrer, B. Alen, Y. Gonzalez, P. Alonso-Gonzalez, D. Fuster, L. Gonzalez, J. Martinez-Pastor, and F. Briones, ACS Nano \textbf{3}, 1513 (2009).
\bibitem{Tommila2012} J. Tommila , C. Strelow, A. Schramm, T.V. Hakkarainen, M. Dumitrescu, T. Kipp, and M. Guina, Nanosc. Res. Lett. \textbf{7}, 313 (2012).
\bibitem{CanetFerrer2013} J. Canet-Ferrer, G. Munoz-Matutano, J. Herranz, D. Rivas, B. Alen, Y. Gonzalez, D. Fuster, L. Gonzalez, and J. Martinez-Pastor, Appl. Phys. Lett. \textbf{103}, 183112 (2013).
\bibitem{Huggenberger2011b} A. Huggenberger, C.Schneider, C.Drescher, S.Heckelmann, T.Heindel, S. Reitzenstein, M. Kamp, S. H\"{o}fling, L. Worschech, A. Forchel, J. Cryst. Growth \textbf{323}, 194 (2011).
\bibitem{Tatebayashi2000} J. Tatebayashi, M. Nishioka, T. Someya, and Y. Arakawa, Appl. Phys. Lett. \textbf{77}, 3382 (2000).
\bibitem{Houel2012} J. Houel, A.V. Kuhlmann, L. Greuter, F. Xue, M. Poggio, B.D. Gerardot, P.A. Dalgarno, A. Badolato, P.M. Petroff, A. Ludwig, D. Reuter, A. D. Wieck, and R.J. Warburton, Phys. Rev. Lett. \textbf{108}, 107401 (2012).
\bibitem{Joens2013} K.D. J\"{o}ns, P. Atkinson, M. M\"{u}ller, M. Heldmaier, S.M. Ulrich, O.G. Schmidt, and P. Michler, Nano Lett. \textbf{13}, 126 (2013).
\bibitem{Schneider2012} C. Schneider, A. Huggenberger, M. Gschrey, P. Gold, S. Rodt, A. Forchel, S. Reitzenstein, S. H\"{o}fling, and M. Kamp, Phys. Stat. Sol. A \textbf{209}, 2379 (2012).
\bibitem{Mohan2010} A. Mohan, P. Gallo, M. Felici, B. Dwir, A. Rudra, J. Faist, and E. Kapon, small \textbf{6}, 1268 (2010).
\bibitem{Schneider2008} C. Schneider, M. Strau\ss, T. S\"{u}nner, A. Huggenberger, D. Wiener, S. Reitzenstein, M. Kamp, S. H\"{o}fling, and A.
Forchel, Appl. Phys. Lett. \textbf{92}, 183101 (2008).
\bibitem{Huo2014} Y.H. Huo,	B.J. Witek,	S. Kumar,	J.R. Cardenas,	J.X. Zhang,	N. Akopian,	R. Singh,	E. Zallo, R. Grifone,	D. Kriegner,	R. Trotta,	F. Ding,	J. Stangl,	V. Zwiller,	G. Bester,	A. Rastelli,	and O.G. Schmidt, Nat. Phys. \textbf{10}, 46 (2014).
\bibitem{Abbarchi2010} M. Abbarchi, T. Kuroda, T. Mano, K. Sakoda, C.A. Mastrandrea, A. Vinattieri, M. Gurioli, and T. Tsuchiya, Phys. Rev. B \textbf{82}, 201301(R) (2010).
\bibitem{Roszak2007} K. Roszak, V.M. Axt, T. Kuhn, and P. Machnikowski, Phys. Rev. B \textbf{76}, 195324 (2007).
\bibitem{Johansen2010} J. Johansen, B. Julsgaard, S. Stobbe, J.M. Hvam, and P. Lodahl, Phys. Rev. B \textbf{81}, 081304(R) (2010).
\bibitem{Seguin2005} R. Seguin, A. Schliwa, S. Rodt, K. P\"{o}tschke, U.W. Pohl, and D. Bimberg, Phys. Rev. Lett. \textbf{95}, 257402 (2005).
\bibitem{Bennett2010} A.J. Bennett,	M.A. Pooley,	R.M. Stevenson,	M.B. Ward,	R.B. Patel,	A. Boyer de la Giroday,	N. Sk\"{o}ld,	I. Farrer,	C.A. Nicoll,	D.A. Ritchie, and A.J. Shields, Nat. Phys. \textbf{6}, 947 (2010).
\bibitem{Bayer2002} M. Bayer, G. Ortner, O. Stern, A. Kuther, A.A. Gorbunov, A. Forchel, P. Hawrylak, S. Fafard, K. Hinzer, T.L. Reinecke, S.N. Walck, J.P. Reithmaier, F. Klopf, and F. Sch\"{a}fer, Phys. Rev. B \textbf{65}, 195315 (2002).
\bibitem{Poem2010} E. Poem,	Y. Kodriano,	C. Tradonsky,	N.H. Lindner,	B.D. Gerardot,	P.M. Petroff,	and D. Gershoni, Nat. Phys. \textbf{6}, 993 (2010).
\bibitem{Benson2000} O. Benson, C. Santori, M. Pelton, and Y. Yamamoto, Phys. Rev. Lett. \textbf{84}, 2513 (2000).
\bibitem{Stevenson2006} R.M. Stevenson, R.J. Young, P. Atkinson, K. Cooper, D.A. Ritchie, and A.J. Shields, Nature \textbf{439}, 179 (2006).
\bibitem{Juska2013} G. Juska,	V. Dimastrodonato,	L.O. Mereni,	A. Gocalinska,	and E. Pelucchi, Nat. Photon. \textbf{7}, 527 (2013).
\bibitem{Kuroda2013} T. Kuroda, T. Mano, N. Ha, H. Nakajima, H. Kumano, B. Urbaszek, M. Jo, M. Abbarchi, Y. Sakuma, K. Sakoda, I. Suemune, X. Marie, and T. Amand, Phys. Rev. B \textbf{88}, 041306(R) (2013).
\bibitem{Larque2008} M. Larqu\'{e}, I. Robert-Philip, and A. Beveratos, Phys. Rev. A \textbf{77}, 042118 (2008).
\bibitem{Young2005} R.J. Young, R.M. Stevenson, A.J. Shields, P. Atkinson, K. Cooper, D.A. Ritchie, K.M. Groom, A.I. Tartakovskii, and M.S. Skolnick, Phys. Rev. B \textbf{72}, 113305 (2005).
\bibitem{Kowalik2005} K. Kowalik, O. Krebs, A. Lema\^{i}tre, S. Laurent, P. Senellart, P. Voisin and J.A. Gaj, Appl. Phys. Lett. \textbf{86}, 041907 (2005).
\bibitem{Trotta2012} R. Trotta, E. Zallo, C. Ortix, P. Atkinson, J.D. Plumhof, J. van den Brink, A. Rastelli, and O. G. Schmidt, Phys. Rev. Lett. \textbf{109}, 147401 (2012).
\bibitem{Shields2007} A.J. Shields, Nat. Photon. \textbf{1}, 215 (2007).
\bibitem{Borri2001} P. Borri, W. Langbein, S. Schneider, U. Woggon, R.L. Sellin, D. Ouyang, and D. Bimberg, Phys. Rev. Lett. \textbf{87}, 157401 (2001).
\bibitem{Muljarov2004} E.A. Muljarov and R. Zimmermann, Phys. Rev. Lett. \textbf{93}, 237401 (2004).
\bibitem{Ramsay2010b} A.J. Ramsay, A.V. Gopal, E.M. Gauger, A. Nazir, B.W. Lovett, A.M. Fox, and M. S. Skolnick, Phys. Rev. Lett. \textbf{104}, 017402 (2010).
\bibitem{Madsen2013} K.H. Madsen, P. Kaer, A. Kreiner-M\o ller, S. Stobbe, A. Nysteen, J. M\o rk, and P. Lodahl, Phys. Rev. B \textbf{88}, 045316 (2013).
\bibitem{Kaer2013} P. Kaer, N. Gregersen, and J. M\o rk, New J. Phys. \textbf{15}, 035027 (2013).
\bibitem{Kaer2014} P. Kaer and J. M\o rk, Phys. Rev. B \textbf{90}, 035312 (2014).
\bibitem{Ates2009} S. Ates, S.M. Ulrich, S. Reitzenstein, A. L\"{o}ffler, A. Forchel, and P. Michler, Phys. Rev. Lett. \textbf{103}, 167402 (2009).
\bibitem{Flagg2012} E.B. Flagg, S.V. Polyakov, T. Thomay, and G.S. Solomon, Phys. Rev. Lett. {\bf 109}, 163601 (2012).
\bibitem{Gazzano2013} O. Gazzano , S. Michaelis de Vasconcellos, C. Arnold, A. Nowak, E. Galopin, I. Sagnes, L. Lanco, A. Lema\^{i}tre, and P. Senellart, Nat. Commun. {\bf 4}, 1425 (2013).
\bibitem{Huber2015} T. Huber, A. Predojevi\'{c}, D. F\"{o}ger, G. Solomon, and G. Weihs, New J. Phys. {\bf 17}, 123025 (2015).
\bibitem{Unsleber2015} S. Unsleber, D.P.S. McCutcheon, M. Dambach, M. Lermer, N. Gregersen, S. H\"{o}fling, J. M\o rk, C. Schneider, and M. Kamp, Phys. Rev. B \textbf{91}, 075413 (2015).
\bibitem{Kuhlmann2013} A.V. Kuhlmann,	J. Houel,	A. Ludwig,	L. Greuter,	D. Reuter,	A.D. Wieck,	M. Poggio, and R.J. Warburton, Nat. Phys. \textbf{9}, 570 (2013).
\bibitem{Konthasinghe2012b} K. Konthasinghe, J. Walker, M. Peiris, C.K. Shih, Y. Yu, M.F. Li, J.F. He, L.J. Wang, H.Q. Ni, Z.C. Niu, and A. Muller, Phys. Rev. B \textbf{85}, 235315 (2012).
\bibitem{He2013b} Y.-M. He, Y. He, Y.-J. Wei, D. Wu, M. Atat\"{u}re, C. Schneider, S. H\"{o}fling, M. Kamp, C.-Y. Lu, and J.-W. Pan, Nat. Nanotechnol. \textbf{8}, 213 (2013).
\bibitem{Gao2013} W.B. Gao, P. Fallahi, E. Togan, A. Delteil, Y.S. Chin, J. Miguel-Sanchez, and A. Imamoglu, Nat. Commun. \textbf{4}, 2744 (2013).
\bibitem{Gold2014} P. Gold, A. Thoma, S. Maier, S. Reitzenstein, C. Schneider, S. H\"{o}fling, and M. Kamp, Phys. Rev. B \textbf{89}, 035313 (2014).
\bibitem{Rodt2005} S. Rodt, A. Schliwa, K. P\"{o}tschke, F. Guffarth, and D. Bimberg, Phys. Rev. B \textbf{71}, 155325 (2005).
\bibitem{Ding2016} X. Ding, Y. He, Z.-C. Duan, N. Gregersen, M.-C. Chen, S. Unsleber, S. Maier, C. Schneider, M. Kamp, S H\"{o}fling, C.-Y. Lu, and J.-W. Pan, Phys. Rev. Lett. {\bf 116} 020401 (2016).
\bibitem{Somaschi2016} N. Somaschi, V. Giesz, L. De Santis, J.C. Loredo, M.P. Almeida, G. Hornecker, S.L. Portalupi, T. Grange, C. Ant\'{o}n, J. Demory, C. G\'{o}mez, I. Sagnes, N.D. Lanzillotti-Kimura, A. Lema\'{i}tre, A. Auffeves, A.G. White, L. Lanco, and P. Senellart, Nat. Photon. {\bf 10}, 340 (2016).
\bibitem{Wei2014} Y.-J. Wei, Y.-M. He, M.-C. Chen, Y.-N. Hu, Y. He, D. Wu, C. Schneider, M. Kamp, S. H\"{o}fling, C.-Y. Lu, and J.-W. Pan, Nano Lett. \textbf{14}, 6515 (2014).
\bibitem{Birkedahl2001} D. Birkedal, K. Leosson, and J.M. Hvam, Phys. Rev. Lett. \textbf{87}, 227401 (2001).
\bibitem{Muller2007} A. Muller, E.B. Flagg, P. Bianucci, X.Y. Wang, D.G. Deppe, W. Ma, J. Zhang, G.J. Salamo, M. Xiao, and C.K. Shih, Phys. Rev. Lett. \textbf{99}, 187402 (2007).
\bibitem{Flagg2009} E.B. Flagg, A. Muller, J.W. Robertson, S. Founta, D.G. Deppe, M. Xiao, W. Ma, G.J. Salamo, and C.K. Shih, Nat. Phys. \textbf{5}, 203 (2009).
\bibitem{Vamivakas2009} A.N. Vamivakas, Y. Zhao, C.-Y. Lu, and M. Atat\"{u}re, Nat. Phys. \textbf{5}, 198 (2009).
\bibitem{Moelbjerg2012} A. Moelbjerg, P. Kaer, M. Lorke, and J. M\o rk, Phys. Rev. Lett. \textbf{108}, 017401 (2012).
\bibitem{Ulrich2011} S.M. Ulrich, S. Ates, S. Reitzenstein, A. L\"{o}ffler, A. Forchel, and P. Michler, Phys. Rev. Lett. \textbf{106}, 247402 (2011).
\bibitem{He2013a} Y. He, Y.-M. He, Y.-J. Wei, X. Jiang, M.-C. Chen, F.-L. Xiong, Y. Zhao, C. Schneider, M. Kamp, S. H\"{o}fling, C.-Y. Lu, and J.-W. Pan, Phys. Rev. Lett. \textbf{111}, 237403 (2013).
\bibitem{Peiris2014} M. Peiris, K. Konthasinghe, Y. Yu, Z.C. Niu, and A. Muller, Phys. Rev. B \textbf{89}, 155305 (2014).
\bibitem{Lu2010} C.-Y. Lu, Y. Zhao, A.N. Vamivakas, C. Matthiesen, S. F\"{a}lt, A. Badolato, and M. Atat\"{u}re, Phys. Rev.B \textbf{81}, 035332 (2010).
\bibitem{Xu2007} X.D. Xu, B. Sun, P.R. Berman, D.G. Steel, A.S. Bracker, D. Gammon, and L.J. Sham, Science \textbf{317}, 929 (2007).
\bibitem{Mollow1969} B.R. Mollow, Phys. Rev. \textbf{188}, 1969 (1969).
\bibitem{He2015} Y. He, Y.-M. He, J. Liu, Y.-J. Wei, H. Y. Ramírez, M. Atat\"{u}re, C. Schneider, M. Kamp, S. H\"{u}fling, C.-Y. Lu, and J.-W. Pan, Phys. Rev. Lett. \textbf{114}, 097402 (2015).
\bibitem{Kamada2001} H. Kamada, H. Gotoh, J. Temmyo, T. Takagahara, and H. Ando, Phys. Rev. Lett. \textbf{87}, 246401 (2001).
\bibitem{Stievater2001} T.H. Stievater, X.Q. Li, D.G. Steel, D. Gammon, D.S. Katzer, D. Park, C. Piermarocchi, and L.J. Sham, Phys. Rev. Lett. \textbf{87}, 133603 (2001).
\bibitem{Matthiesen2013} C. Matthiesen,	M. Geller,	C.H.H. Schulte,	C. Le Gall,	J. Hansom, Z.Y. Li,	M. Hugues, E. Clarke, and M. Atat\"{u}re, Nat. Comm. \textbf{4}, 1600 (2013).
\bibitem{Matthiesen2012} C. Matthiesen, A.N. Vamivakas, and M. Atat\"{u}re, Phys. Rev. Lett. \textbf{108}, 093602 (2012).
\bibitem{Nguyen2011} H.S. Nguyen, G. Sallen, C. Voisin, Ph. Roussignol, C. Diederichs, and G. Cassabois, Appl. Phys. Lett. \textbf{99}, 261904 (2011).
\bibitem{McCutcheon2013} D.P.S. McCutcheon and A. Nazir, Phys. Rev. Lett. \textbf{110}, 217401 (2013).
\bibitem{Conterio2013} M.J. Conterio, N. Sk\"{o}ld, D.J.P. Ellis, I. Farrer, D.A. Ritchie, and A.J. Shields, Appl. Phys. Lett. \textbf{103}, 162108 (2013).
\bibitem{Melet2008} R. Melet, V. Voliotis, A. Enderlin, D. Roditchev, X.L. Wang, T. Guillet, and R. Grousson, Phys. Rev. B \textbf{78}, 073301 (2008).
\bibitem{Jayakumar2013} H. Jayakumar, A. Predojevi\'{c}, T. Huber, T. Kauten, G.S. Solomon, and G. Weihs, Phys. Rev. Lett. {\bf 110}, 135505 (2013).
\bibitem{Makhonin2014} M.N. Makhonin, J.E. Dixon, R.J. Coles, B. Royall, I.J. Luxmoore, E. Clarke, M. Hugues, M.S. Skolnick, and A.M. Fox, Nano Lett. \textbf{14}, 6997 (2014).
\bibitem{Reithmaier2015} G. Reithmaier, M. Kaniber, F. Flassig, S. Lichtmannecker, K. M\"{u}ller, A. Andrejew, J. Vu\v{c}kovi\'{c}, R. Gross, and J.J. Finley, Nano Lett. \textbf{15}, 5208 (2015).
\bibitem{Faraon2008b} A. Faraon, I. Fushman, D. Englund, N. Stoltz, P. Petroff, and J. Vu\v{c}kovi\'{c},, Nat. Phys. \textbf{4}, 859 (2008).
\bibitem{Reinhard2012} A. Reinhard, T. Volz, M. Winger, A. Badolato, K. J. Hennessy, E. L. Hu, and A. Imamoglu, Nat. Photon. \textbf{6}, 93 (2012).
\bibitem{Konthasinghe2012a} K. Konthasinghe, M. Peiris, Y. Yu, M.F. Li, J.F. He, L.J. Wang, H.Q. Ni, Z.C. Niu, C.K. Shih, and A. Muller, Phys. Rev. Lett. \textbf{109}, 267402 (2012).
\bibitem{Bylander2003} J. Bylander, I. Robert-Philip, I. Abram, Europ. Phys. Journ. D \textbf{22}, 295 (2003).
\bibitem{Hong1987} C.K. Hong, Z.Y. Ou, and L. Mandel, Phys. Rev. Lett. \textbf{59}, 2044 (1987).
\bibitem{Patel2008} R.B. Patel, A.J. Bennett, K. Cooper, P. Atkinson, C.A. Nicoll, D.A. Ritchie, and A.J. Shields, Phys. Rev. Lett. \textbf{100}, 207405 (2008).
\bibitem{Bennett2009} A.J. Bennett, R.B. Patel, C.A. Nicoll, D.A. Ritchie, and A.J. Shields, Nat. Phys. \textbf{5}, 715 (2009).
\bibitem{Polyakov2011} S.V. Polyakov, A. Muller, E.B. Flagg, A. Ling, N. Borjemscaia, E. Van Keuren, A. Migdall, and G.S. Solomon, Phys. Rev. Lett. \textbf{107}, 157402 (2011).
\bibitem{Konthasinghe2014} K. Konthasinghe, M. Peiris, and A. Muller, Phys. Rev. A \textbf{90}, 023810 (2014).
\bibitem{Ates2012} S. Ates, I. Agha, A. Gulinatti, I. Rech, M.T. Rakher, A. Badolato, and K. Srinivasan, Phys. Rev. Lett. \textbf{109}, 147405 (2012).
\bibitem{Purcell1946} E. Purcell, Phys. Rev. \textbf{69}, 681 (1946).
\bibitem{Kleppner1981} D. Kleppner, Phys. Rev. Lett. \textbf{47}, 233 (1981).
%\bibitem{Mariani2013} S. Mariani, A. Andronico, O. Mauguin, A. Lema\^{i}tre, I. Favero, S. Ducci, and G. Leo, Opt. Lett. \textbf{38}, 3965 (2013).
\bibitem{Balram2014} K.C. Balram, M. Davan\c{c}o, J.Y. Lim, J.D. Song, and K. Srinivasan, Optica \textbf{1}, 414 (2014). 
\bibitem{Luxmoore2013c} I.J. Luxmoore,	R. Toro,	O. Del Pozo-Zamudio,	N.A. Wasley,	E.A. Chekhovich,	A.M. Sanchez,	R. Beanland,	A.M. Fox,	M.S. Skolnick,	H.Y. Liu, and A.I. Tartakovskii, Sci. Rep. \textbf{3}, 1239 (2012).
\bibitem{Tandaechanurat2011} A. Tandaechanurat, S. Ishida, D. Guimard, M. Nomura, S. Iwamoto, Y. Arakawa, Nat. Photon. \textbf{5}, 91 (2011).
\bibitem{Srinivasan2007} K. Srinivasan and O. Painter, Nature \textbf{450}, 862 (2007).
\bibitem{Schneider2015} C. Schneider, P. Gold, S. Reitzenstein, S. H\"{o}fling, and M. Kamp, arXiv:1510.05447 (2015).
\bibitem{Lermer2012} M. Lermer, N. Gregersen, F. Dunzer, S. Reitzenstein, S. H\"{o}fling, J. M\o rk, L. Worschech, M. Kamp, and A. Forchel, Phys. Rev. Lett. \textbf{108}, 057402 (2012).
\bibitem{Joannopoulos2008} J.D. Joannopoulos, S.G. Johnson, J.N. Winn, and R.D. Meade, \textit{Photonic Crystals: Molding the Flow of Light}, 2nd Ed. (Princeton University Press, 2008).
\bibitem{Nozaki2010} K. Nozaki, T. Tanabe, A. Shinya, S. Matsuo, T. Sato, H. Taniyama, and M. Notomi,  Nat. Photon. \textbf{4}, 477 (2010).
\bibitem{Englund2005} D. Englund, D. Fattal, E. Waks, G. Solomon, B.Y. Zhang, T. Nakaoka, Y. Arakawa, Y. Yamamoto, and J. Vu\v{c}kovi\'{c}, Phys. Rev. Lett. \textbf{95}, 013904  (2005).
\bibitem{Shirane2007} M. Shirane, S. Kono, J. Ushida, S. Ohkouchi, N. Ikeda, Y. Sugimoto, and A. Tomita, J. Appl. Phys. \textbf{101}, 073107 (2007).
\bibitem{Fan2010} W.J. Fan, Z.B. Hao, E. Stock, J.B. Kang, Y. Luo, and D. Bimberg, Semicond. Sci. Technol. \textbf{26}, 014014 (2011).
\bibitem{Saucer2013} T.W. Saucer and V. Sih, Opt. Expr. \textbf{21}, 20831 (2013).
\bibitem{Akahane2003} Y. Akahane, T. Asano, B.-S. Song, and S. Noda, Nature \textbf{425}, 944 (2003).
\bibitem{Akahane2005} Y. Akahane, T. Asano, B.-S. Song, and S. Noda, Opt. Expr. \textbf{13}, 1202 (2005).
\bibitem{Combrie2008} S. Combri\'{e}, A. De Rossi, Q.V. Tran, and H. Benisty, Opt. Lett. \textbf{33}, 1908 (2008).
\bibitem{Minkov2014} M. Minkov and V. Savona, Sci. Rep. \textbf{4}, 5124 (2014).
\bibitem{Takagi2012} H. Takagi, Y. Ota, N. Kumagai, S. Ishida, S. Iwamoto, and Y. Arakawa, Opt. Expr. \textbf{20}, 28292 (2012).
\bibitem{Enderlin2012} A. Enderlin, Y. Ota, R. Ohta, N. Kumagai, S. Ishida, S. Iwamoto, and Y. Arakawa, Phys. Rev. B \textbf{86}, 075314 (2012).
\bibitem{ViasnoffSchwoob2005} E. Viasnoff-Schwoob, C. Weisbuch, H. Benisty, S. Olivier, S. Varoutsis, I. Robert-Philip, R. Houdre, and C.J.M. Smith, Phys. Rev. Lett. \textbf{95}, 183901 (2005).
\bibitem{Lecamp2007} G. Lecamp, P. Lalanne, and J.P. Hugonin, Phys. Rev. Lett. \textbf{99}, 023902 (2007).
\bibitem{MangaRao2007a} V.S.C. Manga Rao and S. Hughes, Phys. Rev. B \textbf{75}, 205437 (2007).
\bibitem{MangaRao2007b} V.S.C. Manga Rao and S. Hughes, Phys. Rev. Lett. \textbf{99}, 193901 (2007).
\bibitem{Dewhurst2010} S.J. Dewhurst, D. Granados, D.J. Ellis, A.J. Bennett, R.B. Patel, I. Farrer, D. Anderson, G.A. Jones, D.A. Ritchie, and A.J. Shields, Appl. Phys. Lett. \textbf{96}, 031109 (2010).
\bibitem{Geveaux2006} D.G. Gevaux, A.J. Bennett, R.M. Stevenson, A.J. Shields, P. Atkinson, J. Griffiths, D. Anderson, G.A.C. Jones, and D.A. Ritchie, Appl. Phys. Lett. \textbf{88}, 131101 (2006).
\bibitem{Faraon2007} A. Faraon, D. Englund, I. Fushman, J. Vu\v{c}kovi\'{c}, N. Stoltz, and P. Petroff, Appl. Phys. Lett. \textbf{90}, 213110 (2007).
\bibitem{Faraon2009} A. Faraon and J. Vu\v{c}kovi\'{c}, Appl. Phys. Lett. \textbf{95}, 043102 (2009).
\bibitem{Kim2011} H.C. Kim, T.C. Shen, D. Sridharan, G.S. Solomon, and E. Waks, Appl. Phys. Lett. \textbf{98}, 091102 (2011).
\bibitem{Beetz2013} J. Beetz, T. Braun, C. Schneider, S. H\"{o}fling, and M. Kamp, Semicond. Sci. Technol. \textbf{28}, 122002 (2013).
\bibitem{Sun2013} S. Sun, H.C. Kim, G.S. Solomon, and E. Waks, Appl. Phys. Lett. \textbf{103}, 151102 (2013).
\bibitem{Fry2000} P.W. Fry, I.E. Itskevich, D.J. Mowbray, M.S. Skolnick, J.J. Finley, J.A. Barker, E.P. O'Reilly, L.R. Wilson, I.A. Larkin, P.A. Maksym, M. Hopkinson, M. Al-Khafaji, J.P.R. David, A.G. Cullis, G. Hill, and J.C. Clark, Phys. Rev. Lett. \textbf{84}, 733 (2000).
\bibitem{Hofbauer2007} F. Hofbauer, S. Grimminger, J. Angele, G. B\"{o}hm, R. Meyer, M.-C. Amann, and J.J. Finley, Appl. Phys. Lett. \textbf{91}, 201111 (2007).
\bibitem{Chauvin2009} N. Chauvin, C. Zinoni, M. Francardi, A. Gerardino, L. Balet, B. Alloing, L.H. Li, and A. Fiore, Phys. Rev. B \textbf{80}, 241306 (2009).
\bibitem{Laucht2009} A. Laucht, F. Hofbauer, N. Hauke, J. Angele, S. Stobbe, M. Kaniber, G. B\"{o}hm, P. Lodahl, M.-C. Amann, and J.J. Finley, New J. Phys. \textbf{11}, 023034 (2009).
\bibitem{Thon2011} S.M. Thon, H. Kim, C. Bonato, J. Gudat, J. Hagemeier, P.M. Petroff, and D. Bouwmeester, Appl. Phys. Lett. \textbf{99}, 161102 (2011).
\bibitem{Pagliano2014} F. Pagliano, Y.J. Cho, T. Xia, F. van Otten, R. Johne, and A. Fiore, Nat. Commun. \textbf{5}, 5786 (2014).
\bibitem{Petruzzella2015} M. Petruzzella, T. Xia, F. Pagliano, S. Birindelli, L. Midolo, Z. Zobenica, L.H. Li, E.H. Linfield, and A. Fiore, Appl. Phys. Lett. \textbf{107}, 141109 (2015).
\bibitem{Faraon2008} A. Faraon, D. Englund, D. Bulla, B. Luther-Davies, B.J. Eggleton, N. Stoltz, P. Petroff, and J. Vu\v{c}kovi\'{c}, Appl. Phys. Lett. \textbf{92}, 043123 (2008).
\bibitem{Hennessy2005} K. Hennessy, A. Badolato, A. Tamboli, P.M. Petroff, E. Hu, M. Atat\"{u}re, J. Dreiser, and A. Imamoglu, Appl. Phys. Lett. \textbf{87}, 021108 (2005).
\bibitem{Hennessy2006} K. Hennessy, C. H\"{o}gerle, E. Hu, A. Badolato, and A. Imamoglu, Appl. Phys. Lett. \textbf{89}, 041118 (2006).
\bibitem{Lee2009} H.S. Lee, S. Kiravittaya, S. Kumar, J.D. Plumhof, L. Balet, L.H. Li, M. Francardi, A. Gerardino, A. Fiore, A. Rastelli, and O.G. Schmidt, Appl. Phys. Lett. \textbf{95}, 191109 (2009).
\bibitem{Intonti2012} F. Intonti, N. Caselli, S. Vignolini, F. Riboli, S. Kumar, A. Rastelli, O.G. Schmidt, M. Francardi, A. Gerardino, L. Balet, L.H. Li, A. Fiore, and M. Gurioli, Appl. Phys. Lett. \textbf{100}, 033116 (2012).
\bibitem{Piggott2014} A.Y. Piggott, K.G. Lagoudakis, T. Sarmiento, M. Bajcsy, G. Shambat, and J. Vu\v{c}kovi\'{c}, Opt. Expr. \textbf{22}, 15017 (2014).
\bibitem{Mosor2005} S. Mosor, J. Hendrickson, B.C. Richards, J. Sweet, G. Khitrova, H.M. Gibbs, T. Yoshie, A. Scherer, O.B. Shchekin, and D.G. Deppe, Appl. Phys. Lett. \textbf{87}, 141105 (2005).
\bibitem{Intonti2009} F. Intonti, S. Vignolini, F. Riboli, M. Zani, D.S. Wiersma, L. Balet, L.H. Li, M. Francardi, A. Gerardino, A. Fiore, and M. Gurioli, Appl. Phys. Lett. \textbf{95}, 173112 (2009).
\bibitem{Vignolini2010} S. Vignolini, F. Riboli, D.S. Wiersma, L. Balet, L.H. Li, M. Francardi, A. Gerardino, A. Fiore, M. Gurioli, and F. Intonti, Appl. Phys. Lett. \textbf{96}, 141114 (2010).
\bibitem{Speijcken2012} N.W.L. Speijcken, M.A. D\"{u}ndar, A.C. Bedoya, C. Monat, C. Grillet, P. Domachuk, R. N\"{o}tzel, B.J. Eggleton, and R.W. van der Heijden, Appl. Phys. Lett. \textbf{100}, 261107 (2012).
\bibitem{Koenderink2005} A.F. Koenderink, M. Kafesaki, B.C. Buchler, and V. Sandoghdar, Phys. Rev. Lett. \textbf{95}, 153904 (2005).
\bibitem{Vignolini2008} S. Vignolini, F. Intonti, L. Balet, M. Zani, F. Riboli, A. Vinattieri, D.S. Wiersma, M. Colocci, L.H. Li, M. Francardi, A. Gerardino, A. Fiore, and M. Gurioli, Appl. Phys. Lett. \textbf{93}, 023124 (2008).
\bibitem{Notomi2006} M. Notomi, H. Taniyama, S. Mitsugi, and E. Kuramochi, Phys. Rev. Lett. \textbf{97}, 023903 (2006).
\bibitem{Midolo2011} L. Midolo, P.J. van Veldhoven, M.A. D\"{u}ndar, R. N\"{o}tzel, and A. Fiore, Appl. Phys. Lett. \textbf{98}, 211120 (2011).
\bibitem{Midolo2012} L. Midolo, F. Pagliano, T.B. Hoang, T. Xia, F.W.M. van Otten, L.H. Li, E. Linfield, M. Lermer, S. H\"{o}fling, and A. Fiore, Appl. Phys. Lett. \textbf{101}, 091106 (2012).
\bibitem{Sugimoto2004} Y. Sugimoto, Y. Tanaka, N. Ikeda, Y. Nakamura, and K. Asakawa, Opt. Expr. \textbf{12}, 1090 (2004).
\bibitem{Fattahpoor2013} S. Fattah poor, T.B. Hoang, L. Midolo, C.P. Dietrich, L.H. Li, E.H. Linfield, J.F.P. Schouwenberg, T. Xia, F.M. Pagliano, F.W.M. Otten, and A. Fiore, Appl. Phys. Lett. \textbf{102}, 131105 (2013).
\bibitem{Stievater2009} T.H. Stievater, D. Park, W.S. Rabinovich, M.W. Pruessner, S. Kanakaraju, C.J.K. Richardson, and J.B. Khurgin, Opt. Expr. \textbf{18}, 885 (2010).
\bibitem{Tatebayashi2007} J. Tatebayashi, R.B. Laghumavarapu, N. Nuntawong, and D.L. Huffaker, Electr. Lett. \textbf{43}, 410 (2007).
\bibitem{Shin2008} J.H. Shin, Y.-C. Chang, and N. Dagli, Appl. Phys. Lett. \textbf{92}, 201103 (2008).
\bibitem{deLima2006} M.M. de Lima Jr., M. Beck, R. Hey, and P.V. Santos, Appl. Phys. Lett. \textbf{89}, 121104 (2006).
\bibitem{Poot2014} M. Poot and H.X. Tang, Appl. Phys. Lett. \textbf{104}, 061101 (2014).
\bibitem{Joens2015} K.D. J\"{o}ns, U. Rengstl, M. Oster, F. Hargart, M. Heldmaier, S. Bounouar, S.M. Ulrich, M. Jetter, and P. Michler, J. Phys. D: Appl. Phys. \textbf{48}, 085101 (2015).
\bibitem{Prtljaga2014} N. Prtljaga, R.J. Coles, J. O'Hara, B. Royall, E. Clarke, A.M. Fox, and M.S. Skolnick, Appl. Phys. Lett. \textbf{104}, 231107 (2014).
\bibitem{Rengstl2015} U. Rengstl, M. Schwartz, T. Herzog, F. Hargart, M. Paul, S.L. Portalupi, M. Jetter, and P. Michler, Appl. Phys. Lett. \textbf{107}, 021101 (2015).
\bibitem{Kasprzak2010} J. Kasprzak, S. Reitzenstein, E.A. Muljarov, C. Kistner, C. Schneider, M. Strauss, S. H\"{o}fling, A. Forchel, and W. Langbein, Nat. Mater. \textbf{9}, 304 (2010).
\bibitem{Loo2012} V. Loo, C. Arnold, O. Gazzano, A. Lema\'{i}tre, I. Sagnes, O. Krebs, P. Voisin, P. Senellart, and L. Lanco, Phys. Rev. Lett. {\bf 109}, 166806 (2012).
\bibitem{Volz2012} T. Volz, A. Reinhard, M. Winger, A. Badolato, K.J. Hennessy, E.L. Hu, and A. Imamoglu, Nat. Photon. \textbf{6}, 609 (2012).
\bibitem{Englund2012} D. Englund, A. Majumdar, M. Bajcsy, A. Faraon, P. Petroff, and J. Vu\v{c}kovi\'{c}, Phys. Rev. Lett. \textbf{108}, 093604 (2012).
\bibitem{Bose2012} R. Bose, D. Sridharan, H. Kim, G.S. Solomon, and E. Waks, Phys. Rev. Lett. \textbf{108}, 227402 (2012).
\bibitem{Warburton2013} R.J. Warburton, Nat. Mater. \textbf{12}, 483 (2013).
\bibitem{Petta2005} J.R. Petta, A.C. Johnson, J.M. Taylor, E.A. Laird, A. Yacoby, M.D. Lukin, C.M. Marcus, M.P. Hanson, A.C. Gossard, Science \textbf{309}, 2180 (2005).
\bibitem{Berezovsky2008} J. Berezovsky, M.H. Mikkelsen, N.G. Stoltz, L.A. Coldren, D.D. Awschalom, Science \textbf{320}, 349 (2008).
\bibitem{Atature2007} M. Atat\"{u}re, J. Dreiser, A. Badolato, and A. Imamoglu, Nat. Phys. \textbf{3}, 101 (2007).
\bibitem{Luxmoore2013a} I.J. Luxmoore, N.A. Wasley, A.J. Ramsay, A.C.T. Thijssen, R. Oulton, M. Hugues, S. Kasture, V.G. Achanta, A.M. Fox, and M.S. Skolnick, Phys. Rev. Lett. \textbf{110}, 037402 (2013).
\bibitem{Luxmoore2013b} I.J. Luxmoore, N.A. Wasley, A.J. Ramsay, A.C.T. Thijssen, R. Oulton, M. Hugues, A.M. Fox, and M.S. Skolnick, Appl. Phys. Lett. \textbf{103}, 241102 (2013).
\bibitem{Young2015} A.B. Young, A.C.T. Thijssen, D.M. Beggs, P. Androvitsaneas, L. Kuipers, J.G. Rarity, S. Hughes, and R. Oulton, Phys. Rev. Lett. \textbf{115}, 153901 (2015).
\bibitem{Soellner2015} I. S\"{o}llner, S. Mahmoodian, S. Lindskov Hansen, L. Midolo, A. Javadi, G. Kir\v{s}anske, T. Pregnolato, H. El-Ella, E.H. Lee, J.D. Song, S. Stobbe, and P. Lodahl, Nat. Nanotechnol. \textbf{10}, 775 (2015).
\bibitem{Gol'tsman2001} G.N. Gol'tsman, O. Okunev, G. Chulkova, A. Lipatov, A. Semenov, K. Smirnov, B. Voronov, A. Dzardanov, C. Williams, and R. Sobolewski, Appl. Phys. Lett. \textbf{79}, 705 (2001).
\bibitem{Hadfield2009} R.H. Hadfield, Nat. Photon. \textbf{3}, 696 (2009).
\bibitem{Natarajan2012} C.M. Natarajan, M.G. Tanner, and R.H. Hadfield, Supercond. Sci. Technol. \textbf{25}, 063001 (2012).
\bibitem{Gaggero2010} A. Gaggero, S. Jahanmiri Nejad, F. Marsili, F. Mattioli, R. Leoni, D. Bitauld, D. Sahin, G.J. Hamhuis, R. N\"{o}tzel, R. Sanjines, and A. Fiore, Appl. Phys. Lett. \textbf{97}, 151108 (2010).
\bibitem{Renema2014} J.J. Renema, R. Gaudio, Q. Wang, Z. Zhou, A. Gaggero, F. Mattioli, R. Leoni, D. Sahin, M.J.A. de Dood, A. Fiore, and M.P. van Exter, Phys. Rev. Lett. \textbf{112}, 117604 (2014). 
\bibitem{Renema2015} J.J. Renema, Q. Wang, R. Gaudio, I. Komen, K. op 't Hoog, D. Sahin, A. Schilling, M.P. van Exter, A. Fiore, A. Engel, and M.J.A. de Dood, Nano Lett. \textbf{15}, 4541 (2015).
\bibitem{Kerman2006} A.J. Kerman, E.A. Dauler, W.E. Keicher, J.K.W. Yang, K.K. Berggren, G.N. Gol'tsman, and B. Voronov, Appl. Phys. Lett. \textbf{88}, 111116 (2006).
\bibitem{Marsili2011} F. Marsili, F. Najafi, E. Dauler, F. Bellei, X.L. Hu, M. Csete, R.J. Molnar, and K.K. Berggren, Nano Lett. \textbf{11}, 2048 (2011).
\bibitem{Gaudio2014} R. Gaudio, K.P.M. op 't Hoog, Z. Zhou, D. Sahin, and A. Fiore, Appl. Phys. Lett. \textbf{105}, 222602 (2014).
\bibitem{Calkins2013} B. Calkins, P.L. Mennea, A.E. Lita, B.J. Metcalf, W.S. Kolthammer, A. Lamas-Linares, J.B. Spring, P.C. Humphreys, R.P. Mirin, J.C. Gates, P.G.R. Smith, I.A. Walmsley, T. Gerrits, and S.W. Nam, Opt. Expr. \textbf{21}, 22657 (2013).
\bibitem{Ferrari2015} S. Ferrari, O. Kahl, V. Kovalyuk, G.N. Gol'tsman, A. Korneev, and W.H.P. Pernice, Appl. Phys. Lett. \textbf{106}, 151101 (2015).
\bibitem{Kahl2015} O. Kahl, S. Ferrari, V. Kovalyuk, G.N. Gol'tsman, A. Korneev, and W.H.P. Pernice, Sci. Rep. \textbf{5}, 10941 (2015).
\bibitem{Najafi2015} F. Najafi, J. Mower, N.C. Harris, F. Bellei, A. Dane, C. Lee, X.L. Hu, P. Kharel, F. Marsili, S. Assefa, K.K. Berggren, and D. Englund, Nat. Commun. \textbf{6}, 5873 (2015).
\bibitem{Baek2011} B. Baek, A.E. Lita, V. Verma, and S.W. Nam, Appl. Phys. Lett. \textbf{98}, 251105 (2011).
\bibitem{Marsili2013} F. Marsili, V.B. Verma, J.A. Stern, S. Harrington, A.E. Lita, T. Gerrits, I. Vayshenker, B. Baek, M.D. Shaw, R.P. Mirin, and S.W. Nam, Nat. Photon. \textbf{7}, 210 (2013).
\bibitem{Sahin2015} D. Sahin, A. Gaggero, J.-W. Weber, I. Agafonov, M.A. Verheijen, F. Mattioli, J. Beetz, M. Kamp, S. H\"{o}fling, M.C.M van de Sanden, R. Leoni, and A. Fiore, IEEE J. Sel. Topics Quantum. Electron. \textbf{21}, 3800210 (2015).
\bibitem{Kerman2007} A.J. Kerman, E.A. Dauler, J.K.W. Yang, K.M. Rosfjord, V. Anant, K.K. Berggren, G.N. Gol'tsman, and B. Voronov, Appl. Phys. Lett. \textbf{90}, 101110 (2007).
\bibitem{Akhlaghi2015} M.K. Akhlaghi,	E. Schelew, J.F. Young, Nat. Commun. \textbf{6}, 8233 (2015).
\bibitem{Sahin2013a} D. Sahin, A. Gaggero, T.B. Hoang, G. Frucci, F. Mattioli, R. Leoni, J. Beetz, M. Lermer, M. Kamp, S. H\"{o}fling, and A. Fiore, Opt. Expr. \textbf{21}, 11162 (2013).
\bibitem{Dauler2009} E.A. Dauler, A.J. Kerman, B.S. Robinson, J.K.W. Yang, G.G.B. Voronovc, S.A. Hamilton, and K.K. Berggren, J. Mod. Opt. \textbf{56}, 364 (2009).
\bibitem{Reithmaier2013a} G. Reithmaier, S. Lichtmannecker, T. Reichert, P. Hasch, K. M\"{u}ller, M. Bichler, R. Gross, and J.J. Finley, Sci. Rep. \textbf{3}, 1901 (2013).
\bibitem{Divochiy2008} A. Divochiy, F. Marsili, D. Bitauld, A. Gaggero, R. Leoni, F. Mattioli, A. Korneev, V. Seleznev, N. Kaurova, O. Minaeva, G.N. Gol'tsman, K.G. Lagoudakis, M. Benkhaoul, F. Levy, and A. Fiore, Nat. Photon. \textbf{2}, 302 (2008).
\bibitem{Jahanmirinejad2012a} S. Jahanmirinejad and A. Fiore, Opt. Expr. \textbf{20}, 5017 (2012).
\bibitem{Jahanmirinejad2012b} S. Jahanmirinejad, G. Frucci, F. Mattioli, D. Sahin, A. Gaggero, R. Leoni, and A. Fiore, Appl. Phys. Lett. \textbf{101}, 072602 (2012).
\bibitem{Zhou2014} Z. Zhou, S. Jahanmirinejad, F. Mattioli, D. Sahin, G. Frucci, A. Gaggero, R. Leoni, and A. Fiore, Opt. Expr. \textbf{22}, 3475 (2014).
\bibitem{Sahin2013b} D. Sahin, A. Gaggero, Z. Zhou, S. Jahanmirinejad, F. Mattioli, R. Leoni, J. Beetz, M. Lermer, M. Kamp, S. H\"{o}fling, and A. Fiore, Appl. Phys. Lett. \textbf{103}, 111116 (2013).
\bibitem{Shadbolt2012} P.J. Shadbolt, M.R. Verde, A. Peruzzo, A. Politi, A. Laing, M. Lobino, J.C.F. Matthews, M.G. Thompson, and J.L. O'Brien, Nat. Photon. \textbf{6}, 45 (2012).
\bibitem{Zinoni2007} C. Zinoni, B. Alloing, L.H. Li, F. Marsili, A. Fiore, L. Lunghi, A. Gerardino, Yu.B. Vakhtomin, K.V. Smirnov, G.N. Gol'tsman, Appl. Phys. Lett. \textbf{91}, 031106 (2007).
\bibitem{Takemoto2010} K. Takemoto, Y. Nambu, T. Miyazawa, K. Wakui, S. Hirose, T. Usuki, M. Takatsu, N. Yokoyama, K. Yoshino, A. Tomita, S. Yorozu, Y. Sakuma, and Y. Arakawa, Appl. Phys. Expr. \textbf{3}, 092802 (2010).
\bibitem{Takesue2013} H. Takesue, N. Matsuda, E. Kuramochi, W.J. Munro, and M. Notomi, Nat. Commun. \textbf{4}, 2725 (2013).
\bibitem{Atkinson2008} P. Atkinson, O.G. Schmidt, S.P. Bremmer, and D.A. Ritchie, Compt. Rend. Phys. \textbf{9}, 788 (2008).
\bibitem{Schneider2009} C. Schneider, A. Huggenberger, T. S\"{u}unner, T. Heindel, M. Strau\ss, S G\"{o}pfert, P. Weinmann, S. Reitzenstein, L. Worschech, M. Kamp, S. H\"{o}fling, and A. Forchel, Nanotechnology \textbf{20}, 434012 (2009).
\bibitem{Pelucchi2007} E. Pelucchi, S. Watanabe, K. Leifer, Q. Zhu, B. Dwir, P. De Los Rios, and E. Kapon, Nano Lett. \textbf{7}, 1282 (2007).
\bibitem{Prechtel2013} J.H. Prechtel, A.V. Kuhlmann, J. Houel, L. Greuter, A. Ludwig, D. Reuter, A.D. Wieck, and R.J. Warburton, Phys. Rev. X \textbf{3}, 041006 (2013).
\bibitem{Alexe2000} M. Alexe, V. Dragoi, M. Reiche, and U. G\"{o}sele, Electr. Lett. \textbf{36}, 677 (2000).
\bibitem{Buckley2012} S. Buckley, K. Rivoire, and J. Vu\v{c}kovi\'{c}, Rep. Prog. Phys. \textbf{75}, 126503 (2012).
\bibitem{Silverstone2013} J.W. Silverstone, D. Bonneau, K. Ohira, N. Suzuki, H. Yoshida, N. Iizuka, M. Ezaki, C.M. Natarajan, M.G. Tanner, R.H. Hadfield, V. Zwiller, G.D. Marshall, J.G. Rarity, J.L. O'Brien, and M.G. Thompson, Nat. Photon. \textbf{8}, 104 (2014).
\bibitem{Midgall2002} A.L. Migdall, D. Branning, and S. Castelletto, Phys. Rev. A \textbf{66}, 053805 (2002).
\bibitem{FrancisJones2015} R.J.A. Francis-Jones and P.J. Mosley, arXiv:1503.06178 (2015).
\bibitem{Wada1999} H. Wada, H. Sasaki and T. Kamijoh, Solid State Electron. \textbf{43}, 1655 (1999).
\end{thebibliography}
\end{document}